\documentclass[12pt,x11names]{article}
\pdfoutput=1
\usepackage{amssymb,amsmath}
\usepackage{epsfig}
\usepackage{epstopdf}
\usepackage{latexsym}
\usepackage{graphicx}
\usepackage{booktabs}
\usepackage{bbm}
\usepackage{enumitem}
\usepackage[numbers,compress]{natbib}
\usepackage{float}
\numberwithin{equation}{section}

\usepackage[T1]{fontenc}

\usepackage[margin=20pt,small]{caption}
\usepackage{subcaption}

\usepackage[toc]{appendix}

\usepackage{overpic}
\usepackage{tikz}
\usetikzlibrary{decorations.pathreplacing}

\usepackage{array}
\usepackage[all]{xy}

\DeclareCaptionType{equ}[Diagram][]

\usepackage{xcolor}
\usepackage{datetime}
\usepackage[
      colorlinks=false,
      linkcolor=darkblue,  
      urlcolor=blue,    
      filecolor=blue,     
      citecolor=red,
linktocpage=true,
      pdfstartview=FitV,
      bookmarksopen=true,
	  hidelinks
      ]{hyperref}

\ifpdf
\fi

\newcommand*{\boxedcolor}{red}
\makeatletter
\renewcommand{\boxed}[1]{\textcolor{\boxedcolor}{%
  \fbox{\normalcolor\m@th$\displaystyle#1$}}}
\makeatother

\definecolor{cardinal}{rgb}{0.6,0,0}
\definecolor{darkgreen}{rgb}{0,0.5,0}
\definecolor{golden}{rgb}{0.92, 0.7, 0}
\definecolor{midnight}{rgb}{0, 0, 0.5}
\definecolor{darkblue}{rgb}{0.2, 0, 0.8}
\newcommand{\Red}{\color{red}}

\newcommand{\DarkGreen}{\color{darkgreen}}

\def\Re{{\rm Re}} \def\Im{{\rm Im}}

\newcommand{\dd}{\mathrm{d}}

\newcommand{\e}{\mathrm{e}}

\newcommand{\f}[2]{\frac{#1}{#2}}
\newcommand{\R}{\mathbf{R}}

\renewcommand{\Re}{\text{Re}}
\renewcommand{\Im}{\text{Im}}

\newcommand{\Tr}{\text{Tr}~}

\newcommand{\SU}{\mathop{\rm SU}}
\newcommand{\SO}{\mathop{\rm SO}}

\newcommand{\SL}{\mathop{\rm SL}}
\newcommand{\U}{\mathop{\rm {}U}}

\newcommand{\so}{\mathfrak{so}}
\newcommand{\su}{\mathfrak{su}}

\newcommand{\uu}{\mathfrak{u}}

\def\Tr{{\rm Tr}\,}

\def\SL{{\rm SL}}

\def\SO{{\rm SO}}

\def\SU{{\rm SU}}

\def\so{\frak{so}}
\def\su{\frak{su}}

\usepackage{amsthm}

\usepackage{enumerate}
\numberwithin{equation}{section}

\usepackage{color}
\definecolor{dark-gray}{gray}{0.20}
\definecolor{gray}{gray}{0.30}
\definecolor{light-gray}{gray}{0.80}
\definecolor{dark-red}{rgb}{0.7,0,0}
\definecolor{dark-green}{rgb}{0.1,0.4,0}
\definecolor{dark-blue}{rgb}{0.3,0.3,0.7}
\definecolor{light-blue}{rgb}{0.8,0.8,1}

\newcommand{\be}{\begin{equation}}
\newcommand{\ee}{\end{equation}}
\newcommand{\bea}{\begin{eqnarray}}
\newcommand{\eea}{\end{eqnarray}}
\newcommand{\Z}{\mathbf{Z}}
\renewcommand{\Re}{\text{Re}}
\renewcommand{\Im}{\text{Im}}

\newcommand{\Ess}{\text{E}_{7(7)}}

\newcommand{\ess}{\mathfrak{e}_{7(7)}}

\usepackage{hyperref}
\hypersetup{
     bookmarks=true,         
     unicode=false,          
     pdftoolbar=true,        
     pdfmenubar=true,        
     pdffitwindow=false,     
     pdfstartview={FitH},    
     pdftitle={My title},    
     pdfauthor={Author},     
     pdfsubject={Subject},   
     pdfcreator={Creator},   
     pdfproducer={Producer}, 
     pdfkeywords={keyword1} {key2} {key3}, 
     pdfnewwindow=true,      
     colorlinks=true,       
     linkcolor=red,          
     citecolor=Green4,        
     filecolor=magenta,      
     urlcolor=DeepSkyBlue3           
}

\begin{document}  

\begin{titlepage}

\medskip
\begin{center} 
{\large \bf  The Holographic Conformal Manifold of 3d $\mathcal{N}=2$ $S$-fold SCFTs}

\bigskip
\bigskip
\bigskip
\bigskip

{\bf Nikolay Bobev,${}^{\mathtt{J}}$ Fri\dh rik Freyr Gautason,${}^{\mathfrak J}$  and Jesse van Muiden${}^{\mathtt J}$  \\ }
\bigskip
\bigskip
\bigskip
\bigskip
${}^{\mathtt{J}}$Instituut voor Theoretische Fysica, KU Leuven \\
Celestijnenlaan 200D, B-3001 Leuven, Belgium
 \vskip 8mm
 ${}^{\mathfrak J}$University of Iceland, Science Institute,\\
 Dunhaga 3, 107 Reykjav\'ik, Iceland\\
 
\bigskip 
\texttt{nikolay.bobev,~jesse.vanmuiden~@kuleuven.be,~ ffg@hi.is} \\
\end{center}

\bigskip
\bigskip
\bigskip
\bigskip
\bigskip
\bigskip

\begin{abstract} 

\noindent We employ a non-compact gauging of four-dimensional maximal supergravity to construct a two-parameter family of AdS$_4$ J-fold solutions preserving $\mathcal{N}=2$ supersymmetry. All solutions preserve $\uu(1) \times \uu(1)$ global symmetry and in special limits we recover the previously known $\su(2) \times \uu(1)$ invariant $\mathcal{N}=2$ and $\su(2) \times \su(2)$ invariant $\mathcal{N}=4$ J-fold solutions. This family of  AdS$_4$ backgrounds can be uplifted to type IIB string theory and is holographically dual to the conformal manifold of a class of three-dimensional $S$-fold SCFTs obtained from the $\mathcal{N}=4$ $T[\U(N)]$ theory of Gaiotto-Witten. We find the spectrum of supergravity excitations of the AdS$_4$ solutions and use it to study how the operator spectrum of the three-dimensional SCFT depends on the exactly marginal couplings.
  
\end{abstract}

\noindent

\end{titlepage}




\section{Introduction and summary of results}
\label{Sec: Introduction}

String theory should contain many non-geometric vacua. Constructing such backgrounds explicitly is in general a hard task. One possible strategy is to study local supergravity solutions arising in the low energy limit of string theory and ``glue'' them together into a globally non-geometric background using string theoretic dualities. A number of such non-geometric Minkowski vacua of string theory were constructed in the literature, however there are relatively few explicitly known AdS non-geometric vacua, see \cite{Plauschinn:2018wbo} for a review. Recently a class of non-geometric AdS vacua of type IIB string theory was constructed in the literature by making use of the non-trivial $\SL(2,\Z)_{\text{IIB}}$ duality group of the theory, see \cite{Inverso:2016eet,Assel:2018vtq,Guarino:2019oct,Bobev:2019jbi,Guarino:2020gfe,Bobev:2020fon,Gautason:2020mgd,Arav:2021tpk}. The backgrounds have the schematic form
\begin{equation}\label{eq:Jfoldintro}
\text{AdS}_4 \times \tilde S^1 \times \hat{S}^5\,,
\end{equation}
where the tilde indicates that there is non-trivial $\SL(2,\Z)_{\text{IIB}}$ monodromy around the circle. The five-dimensional manifold $\hat{S}^5$ is topologically a five sphere with squashed metric threaded by R-R and NS-NS fluxes. These non-geometric backgrounds can be thought of as arising from geometric $\text{AdS}_4 \times \mathbf{R} \times \hat{S}^5$ solutions of type IIB supergravity in which all NS-NS and R-R fields except the dilaton do not depend on $\mathbf{R}$. The dilaton has a linear dependence on the $\mathbf{R}$ direction which can then be compactified into the $\tilde S^1$ in \eqref{eq:Jfoldintro} by appropriate shifts of the dilaton and making use of the $\SL(2,\Z)_{\text{IIB}}$ symmetry of string theory. We refer to this class of string theory backgrounds as J-folds. This moniker is justified by another vantage point that leads to these non-geometric constructions: The backgrounds in \eqref{eq:Jfoldintro} arise as a limit of Janus solutions that are dual to superconformal interfaces in $\mathcal{N}=4$ SYM \cite{Assel:2018vtq,Bobev:2019jbi,Bobev:2020fon,Arav:2021tpk}. This in turn offers the exciting possibility to study the AdS/CFT correspondence for explicit non-geometric string theory vacua.

There are three distinct methods to construct J-fold backgrounds of string theory. One can work directly in IIB supergravity and look for suitable $\text{AdS}_4 \times \mathbf{R} \times \hat{S}^5$ solutions with linear dilaton which then can be compactified using $\SL(2,\Z)_{\text{IIB}}$. In general however this method is often hard to implement in practice. When $\hat{S}^5$ has a small isometry group the supersymmetry variations or equations of motion of IIB supergravity lead to partial differential equations which are hard to solve in general. Nevertheless, explicit J-fold solutions with $\mathcal{N}=0$ or $\mathcal{N}=1$ supersymmetry and large isometry on $\hat{S}^5$ can be constructed in this way, see \cite{Robb:1984uj,Lust:2009mb,Skenderis:2010vz,Bobev:2019jbi}.\footnote{Note that in the backgrounds descibed in \cite{Lust:2009mb,Skenderis:2010vz,Bobev:2019jbi} the $\hat{S}^5$ can be generalized to an arbitrary Sasaki-Einstein manifold.} A complementary strategy to construct J-fold backgrounds is to use the fact that type IIB supergravity admits a consistent truncation to the maximal five-dimensional $\SO(6)$ gauged supergravity. One can then look for $\text{AdS}_4 \times \mathbf{R}$ solutions of the five-dimensional supergravity in which the dilaton has a linear profile and all other scalar fields are constants. These solutions can then be uplifted to type IIB supergravity using the results in \cite{Lee:2014mla,Baguet:2015sma} and one can then proceed to compactify the $\mathbf{R}$ direction into the $\tilde S^1$ in \eqref{eq:Jfoldintro} by using $\SL(2,\Z)_{\text{IIB}}$. The advantage of this method is twofold. First, the problem of finding the $\text{AdS}_4 \times \mathbf{R}$ solution in five dimensions is essentially algebraic even when $\hat{S}^5$ has little or no isometry. Second, the J-fold solutions in five-dimensional supergravity arise as special limits of families of asymptotically AdS$_5$ Janus solutions. This makes it clear that the 3d SCFT dual to the J-fold AdS$_4$ solution should be closely related to the theory residing at a conformal interface in $\mathcal{N}=4$ SYM. Indeed, this strategy was successfully employed in \cite{Bobev:2019jbi,Bobev:2020fon,Gautason:2020mgd}, see also \cite{Arav:2021tpk,Arav:2020obl}, where supersymmetric J-fold backgrounds were constructed as limits of Janus interface solutions. The third method, which will be the one we employ in this work, is to use yet another consistent truncation of type IIB supergravity. As shown in \cite{Inverso:2016eet} type IIB supergravity on $S^5 \times \mathbf{R}$ admits a consistent truncation to four-dimensional maximal gauged supergravity with the non-compact gauge group $[\SO(1,1)\times \SO(6)]\ltimes \mathbf{R}^{12}$ and a dyonic gauging. To construct J-fold backgrounds one then looks for AdS$_4$ vacuum solutions of the four-dimensional supergravity theory which arise as critical points of the potential for the 70 scalar fields in the theory. Finding such AdS$_4$ vacua then reduces to a complicated but purely algebraic problem. Given such an AdS$_4$ vacuum solution it was shown in \cite{Inverso:2016eet,Guarino:2019oct}, using Exceptional Field Theory results from \cite{Hohm:2013uia,Hohm:2014qga}, how to uplift it to type IIB supergravity where one can proceed to use the $\SL(2,\Z)_{\text{IIB}}$ action on the dilaton to obtain a compact J-fold string theory background. We note that it  should be possible to find a direct way to relate the five-dimensional and four-dimensional supergravity constructions described above using a Scherk-Schwarz type reduction \cite{Scherk:1979zr} and it will be very interesting to establish this in detail.

The goal of this paper is to construct and study a large new family of AdS$_4$ $\mathcal{N}=2$ J-fold backgrounds. There are several results in the literature which point to the existence of such a family of J-folds. It was found in \cite{Guarino:2020gfe} that there is family of $\mathcal{N}=2$ J-fold backgrounds, constructed using the maximal four-dimensional supergravity method described above, which are parametrized by a real parameter $\chi$. We refer to this as Family I in the text below. For generic values of $\chi$ the AdS$_4$ background preserves $\mathcal{N}=2$ supersymmetry and a $\uu(1)_F\times \uu(1)_R$ global symmetry. The $\uu(1)_R$ is mapped to the superconformal R-symmetry in the dual 3d $\mathcal{N}=2$ SCFT, while $\uu(1)_F$ is is mapped to a flavor symmetry. For $\chi=0$ there is a symmetry enhancement and the background enjoys $\su(2)_F$ symmetry. The cosmological constant for this family of AdS$_4$ solutions does not depend on $\chi$. This in turns implies that the $S^3$ free energy of the dual family of SCFTs is invariant under changes of $\chi$ which strongly suggests that $\chi$ describes an exactly marginal deformation in the SCFT. Conformal manifolds in 3d $\mathcal{N}=2$ SCFTs are K\"ahler\footnote{This can be shown along the lines of \cite{Asnin:2009xx}, see also \cite{Tachikawa:2005tq,deAlwis:2013jaa}.} which suggests that the real parameter $\chi$ should be complexified and thus there has to be a more general family of J-folds parametrized by (at least) two real parameters. The $\su(2)_F\times \uu(1)_R$ invariant J-fold solutions has also been independently found in \cite{Bobev:2020fon} using five-dimensional gauged supergravity. It was also observed in \cite{Bobev:2020fon} that the $S^3$ free energy of the SCFT dual to the $\su(2)_F\times \uu(1)_R$ J-fold is the same as the $S^3$ free energy of the $\mathcal{N}=4$ $\su(2)\times \su(2)$ invariant J-fold studied in \cite{Inverso:2016eet,Assel:2018vtq,Bobev:2020fon}. It was proposed in \cite{Bobev:2020fon} that theses special $\mathcal{N}=2$ and $\mathcal{N}=4$ SCFTs belong to the same conformal manifold. This implies that there should be a family $\mathcal{N}=2$ of AdS$_4$ J-fold backgrounds that interpolate between the $\mathcal{N}=2$ and $\mathcal{N}=4$ solutions found in \cite{Guarino:2020gfe,Bobev:2020fon} and \cite{Inverso:2016eet,Assel:2018vtq,Bobev:2020fon}.

We indeed find that these expectations are confirmed by constructing an explicit family of J-fold backgrounds described by two real parameters. We first focus on a consistent truncation of the 4d maximal supergravity which is invariant under a $\Z_2^3$ discrete subgroup of the $\SO(1,1)\times \SO(6)$ symmetry of the theory. Only 14 out of the 70 reals scalars in the supergravity theory survive in this truncation. This 14-scalar model is analytically tractable and one can find that it admits two one-parameter families of analytic $\mathcal{N}=2$ AdS$_4$ vacua. The first one, parametrized by the real number $\chi \in (0,\infty)$, is the one found in \cite{Guarino:2020gfe} and dubbed Family I above. The second one, which we call Family II, is a novel family of analytic solutions parametrized by the real number $\varphi \in (0,\infty)$. The solutions at $\chi=0$ and $\varphi=0$ are equivalent and give the $\su(2)_F\times \uu(1)_R$ J-fold of \cite{Guarino:2020gfe,Bobev:2020fon}. At $\varphi=1$ one finds the $\mathcal{N}=4$ $\su(2)\times \su(2)$ invariant J-fold of \cite{Inverso:2016eet,Assel:2018vtq,Bobev:2020fon}. We therefore find that Family II interpolates between the special $\mathcal{N}=2$ and $\mathcal{N}=4$ J-fold solutions and explicitly realizes the proposal made in \cite{Bobev:2020fon}. Given all this it is natural to expect that Family I and Family II are one-dimensional sections of a two-dimensional space of J-folds. We confirm this expectation by constructing this family of solutions. To do this we need to go beyond the 14-scalar truncation described above and include additional scalar fields invariant under the $\uu(1)_F\times \uu(1)_R$ subgroup of the gauge group of the maximal supergravity theory. Our results amount to a new two-parameter family of analytic $\mathcal{N}=2$ AdS$_4$ vacua which are invariant under $\uu(1)_F\times \uu(1)_R$ and smoothly interpolate between the one-parameter Family I and Family II. These solutions should be the supergravity dual of the conformal manifold of the dual 3d $\mathcal{N}=2$ SCFT. In fact, we find that Family I and Family II furnish the boundaries in the two-dimensional space of solutions. In addition to constructing this two-parameter space of J-folds we are also able to calculate explicitly the spectrum of masses of all four-dimensional supergravity fields for all AdS$_4$ solutions. Using the AdS/CFT dictionary we can map these results to find the spectrum of operator dimensions in the dual SCFT. We perform this calculation explicitly and organize the spectrum of SCFT operators into unitary representations of the superconformal algebra. We find a mixture of long, semi-short, and short multiplets for which we compute all quantum numbers. As expected, the conformal dimensions of short and semi-short multiplets do not depend on the exactly marginal deformations. The long multiplets however exhibit an intricate dependence on the two marginal deformation parameters which we describe explicitly. We also observe multiplet rearrangements at the $\mathcal{N}=2$ $\su(2)_F\times \uu(1)_R$ and $\mathcal{N}=4$ $\su(2)\times \su(2)$ J-fold points which nicely match SCFT expectations. In addition, we compute the supergravity kinetic terms for the scalar moduli $(\varphi,\chi)$ which can be interpreted holographically as the Zamolodchikov metric on the SCFT conformal manifold. Our holographic results offer a rare quantitative window into the conformal manifold of strongly coupled SCFTs and we use them to speculate about the possible global structure of the SCFT conformal manifold in Section~\ref{sec:discussion}.

The supergravity discussion above should have a direct counterpart in the dual SCFT description of the J-fold construction. A concrete proposal for the 3d $\mathcal{N}=3$ SCFT dual to the $\mathcal{N}=4$ J-fold was put forward in \cite{Assel:2018vtq}. The SCFT is constructed by taking the Gaiotto-Witten $T[\U(N)]$ theory \cite{Gaiotto:2008ak} and gauging its $\U(N)\times \U(N)$ flavor symmetry using an $\mathcal{N}=4$ vector multiplet and introducing a Chern-Simons term at level $k$ for the gauge field. In the J-fold construction above $N$ is mapped to the number of D3-branes in type IIB string theory and $k$ is related to the integer that specifies the hyperbolic element of $\SL(2,\Z)_{\text{IIB}}$ used to make the $\tilde{S}^1$ in \eqref{eq:Jfoldintro} compact. The $T[\U(N)]$ theory is a 3d $\mathcal{N}=4$ SCFT which has two $\U(N)$ flavor current multiplets that contain complex scalar superconformal primary operators. Each of these complex scalars, usually referred to as moment map operators, transforms in the adjoint representation of $\U(N)$ and we denote them as $\mu_C$ and $\mu_H$ since they are also associated with the Coulomb and Higgs branch of the theory. The gauging of the diagonal $\U(N)$ subgroup of the $\U(N)\times \U(N)$ flavor symmetry then amounts to introducing the following superpotential, see \cite{Assel:2018vtq,Gang:2018huc,Beratto:2020qyk} for more details,  
\begin{equation}\label{eq:WN3}
W_{\rm UV}^{\mathcal{N}=3} = -\frac{k}{4\pi} \Tr(\Phi^2) + \Tr(\Phi(\mu_H +\mu_C))\,.
\end{equation}
Here $\Phi$ is the adjoint complex scalar field in the $\mathcal{N}=4$ vector multiplet and $k$ is the integer Chern-Simons level. The theory with superpotential \eqref{eq:WN3} is defined in terms of $\mathcal{N}=4$ ingredients but preserves only $\mathcal{N}=3$ supersymmetry. It is also not a conformal theory and it was conjectured in \cite{Assel:2018vtq}, see also \cite{Gang:2018huc}, that there is supersymmetry enhancement in the IR where one finds a non-trivial $\mathcal{N}=4$ SCFT. The effective IR superpotential can be obtained by integrating out $\Phi$ in \eqref{eq:WN3} and reads\footnote{The IR superpotentials we write should be viewed as schematic since the SCFT discussed in \cite{Assel:2018vtq} does not admit a known Lagrangian description in the IR.}
\begin{equation}\label{eq:WN4}
W_{\rm IR}^{\mathcal{N}=4} = -\frac{2\pi}{k} \Tr(\mu_H \mu_C)\,.
\end{equation}
To arrive at this result we have made use of the fact that the operators $\mu_{H,C}$ have nilpotent properties and obey $\Tr(\mu_H^2)=\Tr(\mu_C^2)=0$. The SCFT  described by \eqref{eq:WN4} is the field theory dual of the $\mathcal{N}=4$ J-fold background in string theory. Non-trivial evidence for this claim was presented in \cite{Assel:2018vtq} where the $S^3$ partition function of the SCFT at large $N$ was computed by supersymmetric localization and was shown to agree with the bulk AdS$_4$ on-shell action.

This picture suggests a natural supersymmetric marginal deformation of the $\mathcal{N}=4$ SCFT. We can use that the operator $\Tr(\mu_H \mu_C)$ is marginal in the IR and simply change the superpotential in \eqref{eq:WN4} to 
\begin{equation}\label{eq:WN2}
W_{\rm IR}^{\mathcal{N}=2} = -\frac{2\pi}{k} \Tr(\mu_H \mu_C)+\lambda \Tr(\mu_H \mu_C)\,,
\end{equation}
where $\lambda$ is a complex number. An alternative way obtain the superpotential in \eqref{eq:WN2} is to add the mass term $m^2\Phi^2$ to the UV superpotential in \eqref{eq:WN3}. The continuous parameter $m$ breaks the supersymmetry from $\mathcal{N}=3$  to $\mathcal{N}=2$ and modifies the RG flow. Integrating out $\Phi$ from this mass deformed theory will then lead to the superpotential in \eqref{eq:WN2} where the parameter $\lambda$ is related to the UV mass $m$. We conjecture that the conformal manifold of 3d $\mathcal{N}=2$ SCFT corresponding to the superpotential in \eqref{eq:WN2} is the field theory dual to the two-parameter family of AdS$_4$ J-fold backgrounds we construct in this work. Note that this is in harmony with the superconformal index calculations in \cite{Beratto:2020qyk} where it was argued that the 3d $\mathcal{N}=4$ SCFT described by \eqref{eq:WN4} has a one-dimensional complex conformal manifold. We emphasize that the superpotential in \eqref{eq:WN2} should be treated as a schematic representation of the conformal manifold in the neighborhood of the $\mathcal{N}=4$ SCFT and should not be used to draw conclusions about the global structure of the conformal manifold. We note also that in addition to the $\uu(1)_R$ superconformal R-symmetry which acts on $\mu_H$ and $\mu_C$ with charge $+1$, the superpotential \eqref{eq:WN2} enjoys also a $\uu(1)_F$ symmetry which acts with equal and opposite charges on $\mu_H$ and $\mu_C$. This feature is compatible with the $\uu(1)_F\times \uu(1)_R$ symmetry exhibited by the family of J-fold AdS$_4$ backgrounds we construct.

In the next section we start by presenting two families of analytic J-fold solutions in four-dimensional gauged supergravity each of which depend on a real parameter. We then proceed in Section~\ref{sec:18} to find a larger set of J-fold backgrounds that span a two-dimensional space of solutions dual to the conformal manifold of the dual 3d $\mathcal{N}=2$ SCFT. In Section~\ref{sec:spectroscopy} we present explicit results about the mass spectrum of all four-dimensional supergravity fields, show how they are organized into SCFT operator multiplets, and describe how they depend on the position on the conformal manifold. We conclude with a short discussion in Section~\ref{sec:discussion}. In the appendix we collect some useful results about the unitary representations of 3d $\mathcal{N}=2$ and $\mathcal{N}=4$ superconformal algebras.

\medskip

\textit{Note added:} In the final stages of the preparation of the manuscript we learned about \cite{AGRR} in which the authors construct a one-parameter family of AdS$_4$ solutions that connect the previously known $\mathcal{N}=2$ and $\mathcal{N}=4$ J-fold backgrounds using the five-dimensional gauged supergravity truncation in \cite{Bobev:2016nua}. Our results appear to be compatible with those in \cite{AGRR} but are obtained through a different method based on four-dimensional gauged supergravity. The solutions found in \cite{AGRR} may correspond to what we refer to as Family II below.

\section{J-fold solutions in the 14-scalar model}
\label{sec:14}

The workhorse we use to construct the J-fold backgrounds of interest is the $\mathcal{N}=8$ $[\SO(1,1)\times \SO(6)]\ltimes \mathbf{R}^{12}$ gauged supergravity in four dimensions. The full structure of this theory is rather involved, see  \cite{DallAgata:2011aa,Inverso:2016eet,Guarino:2019oct} and \cite{deWit:2005ub,deWit:2007kvg} for further details. We are interested in AdS$_4$ solutions in this theory and therefore need to look for critical points of the scalar potential. Finding such critical points is in general an unwieldy task since the potential is a complicated function of the 70 scalar fields in the supergravity theory which span an $\Ess/\SU(8)$ coset manifold.\footnote{As discussed recently in \cite{Comsa:2019rcz,Bobev:2019dik,Bobev:2020ttg,Krishnan:2020sfg,Bobev:2020qev} one can efficiently use numerical techniques to construct AdS critical points in maximal gauged supergravity theories.} To render the problem tractable by analytic methods we can reduce the number of scalar fields by studying a subspace of the scalar manifold invariant under a subgroup of the symmetry group of the theory.

A relatively simple consistent truncation on the scalar manifold can be found by imposing $\Z_2\times \Z_2\times \Z_2$ global symmetry. This truncation was constructed in \cite{Guarino:2020gfe} and comprises 14 real scalar fields.\footnote{A consistent truncation with the same global symmetry was employed in \cite{Bobev:2020qev,Bobev:2019dik} to find AdS$_4$ vacua in maximal supergravity theories with different gauge groups.} The scalar manifold of this 14-scalar model consists of seven commuting $\SL(2,\R)/\U(1)$. Each $\SL(2,\R)$ is generated by a positive and negative root generator, $\mathfrak{e}$ and $\mathfrak{f}$, and a Cartan generator. A simple way to parametrize the $\SL(2,\R)$'s is to use
\be\label{70bein14scal}
{\cal V}_i = \e^{\sqrt{2}\Re z\,\mathfrak{e}_i}\cdot \e^{\log\Im z\, [\mathfrak{e}_i,\mathfrak{f}_i]}\,,
\ee
where $i=1,\dots,7$ runs over the seven $\SL(2,\R)$'s. Note that the commutator $[\mathfrak{e}_i,\mathfrak{f}_i]$ is a simple way to write the correct Cartan generator belonging to each $\SL(2,\R)$. This consistent truncation of the maximal supergravity theory can be written as an $\mathcal{N}=1$ gauged supergravity coupled to 7 seven chiral multiplets. This theory can be written in terms of the K\"ahler potential
\be\label{eq:Kahlerpot}
K = -\sum_i \log( 2\Im z_i)\,,
\ee
which determines the kinetic terms for the scalar fields, and the holomorphic superpotential
\be\label{eq:W14def}
W= 2g\big(z_1z_5z_6+z_2z_4z_6+z_3z_4z_5+z_1z_4z_7+z_2z_5z_7+z_3z_6z_7\big)+2gc(1-z_4z_5z_6z_7)\,.
\ee
The scalar potential can then be written as
\be\label{eq:V14def}
V = \e^{K}\Big(K^{i\bar{\jmath}}D_iW D_{\bar{\jmath}}\overline{W}-3W\overline{W})\,,
\ee
where $D_i f = \partial_i f + f \partial_i K$ is the K\"ahler covariant derivative on the scalar manifold and $K^{i\bar{\jmath}}$ is the inverse scalar manifold metric obtained from the K\"ahler potential \eqref{eq:Kahlerpot}. The real constant $g$ is the gauge coupling in the supergravity theory and determines the scale of the AdS$_4$ vacuum solutions. The real constant $c$ is related to the type of gauging used in the maximal supergravity theory with $c=0$ corresponding to the ``standard'' electric gauging and $c\neq 0$ yielding a dyonic gauging. When $c\neq 0$ one can fix $c=1$ without loss of generality. We will be interested in the theory with $c\neq 0$ since this model admits an uplift to type IIB supergravity on $\mathbf{R}\times S^5$, see \cite{Inverso:2016eet,Guarino:2019oct}.

It is instructive to study the symmetries of the 14 scalar model. To this end we consider linear maps on the fields $z_i$ that leave the superpotential \eqref{eq:W14def} invariant.\footnote{There is a trivial $\Z_2$ symmetry of the potential \eqref{eq:V14def} that acts by complex conjugation on all $z_i$, however this is clearly not a symmetry of the superpotential. Since we are mostly interested in supersymmetric AdS$_4$ solutions which are critical points of the superpotential we will not discuss this $\Z_2$ further.} We find that there is a discrete group of order 24  generated by the following three actions on the $z_i$
\be
\begin{split}
f:&\quad z_1\to z_2 \to z_1\,,\quad z_4\to z_5 \to z_4\,,\\
g:&\quad z_1\to z_2\to z_3 \to z_1\,,\quad z_5\to z_7 \to z_6\to z_5\,,\\
h:&\quad z_1\to z_3 \to z_1\,,\quad z_4\to z_5\to z_6 \to z_7 \to z_4\,,
\end{split}
\ee
that leave the superpotential in \eqref{eq:W14def} invariant. This order 24 group is simply the symmetric group $S_4$ defined as
\be\label{eq:S4symdef}
S_4 = \big\langle f,g,h| f^2 = g^3 = h^4 = fgh =e\big\rangle\,.
\ee
Remarkably this is exactly the same symmetry group enjoyed by a 10-scalar consistent truncation of the maximal $\SO(6)$ gauged supergravity in five dimensions \cite{Bobev:2016nua,Bobev:2020ttg}. This, together with the fact that there are a number of AdS$_4$ J-fold solutions in the five-dimensional 10-scalar model \cite{Bobev:2020fon,Gautason:2020mgd,Arav:2021tpk,Arav:2020obl}, suggests a non-trivial relation between the two supergravity consistent truncations which will be interesting to explore further. 

\subsection{AdS$_4$ solutions}

Some AdS$_4$ critical points of the 14 scalar model were studied in \cite{Guarino:2020gfe}. They found a family of ${\cal N}=1$ solutions and a one parameter family of ${\cal N}=2$ solutions. In addition, they found a large family of non-supersymmetric solutions as well as a single ${\cal N}=4$ point. Some of these solutions were independently found in five-dimensional supergravity in \cite{Bobev:2020fon} where it was noticed that the cosmological constant of the ${\cal N}=4$ and the $\su(2)\times \uu(1)$ invariant ${\cal N}=2$ solution was exactly the same. As discussed in Section~\ref{Sec: Introduction}, it was speculated in \cite{Bobev:2020fon} that these two solutions lie on the same conformal manifold. Motivated by these results we revisit the 14-scalar model described above and look for new $\mathcal{N}=2$ AdS$_4$ vacua. These solutions should be critical points of the holomorphic superpotential in \eqref{eq:W14def}, i.e. we need to solve the system of algebraic equations
\begin{equation}
D_i W = 0\,.
\end{equation}
We find two distinct families of solutions. Both of them are invariant under a $\mathbf{Z}_2$ subgroup of the $S_4$ symmetry group of the 14-scalar model.

The first family of solutions, which we call Family I, is the one found in \cite{Guarino:2020gfe} and recently analyzed further in \cite{Giambrone:2021zvp,Guarino:2021kyp}. It is invariant under the discrete action $\Z_2=\langle h^2 | h^4=e\rangle$. The 7 complex scalar fields take the following constant values
\be\label{eq:FamIzs}
\text{Family I}:\quad {\bf z} = \left( -c\chi
   +\frac{ic}{\sqrt{2}},i c,
   c\chi+\frac{ic}{\sqrt{2}},i,\frac{1+i}{\sqrt{2}},i,\frac{1+i}{\sqrt{2}}\right)\,.
\ee
The real parameter $\chi \in [0,\infty)$ can be thought of as one of the 14 real scalars in this model. It has vanishing mass and acts as a modulus and a natural coordinate that parametrizes the space of solutions. The scalar potential for the family of solutions \eqref{eq:FamIzs} takes the value $V= -3 g^2/c$. The complete mass spectrum of all four-dimensional supergravity fluctuations can be computed for all values of $\chi$ using the formulae presented in \cite{Gallerati:2014xra}. We will discuss this spectrum in detail in Section~\ref{sec:spectroscopy}. Here we limit ourselves to presenting the gravitino masses which are
\be\label{eq:gravmassesFamI}
m^2_{3/2} L^2 ~:~ 1_2\,,\quad 4_2 \,,\quad ( 2+\chi^2)_4\,,
\ee
where $L$ is the AdS$_4$ radius and the subscript denotes the degeneracy of each mass and the masses have been normalized in such a way that $m^2L^2 =1$ denotes an unbroken supersymmetry. This family of solutions preserves $\uu(1)\times\uu(1)$ continuous symmetry which can be deduced from the two massless vector fields present in the supergravity spectrum. One of the $\uu(1)$'s has the interpretation as the superconformal R-symmetry of the dual conformal field theory while the other is a flavor symmetry. At $\chi=0$ there are two more massless vector fields in the spectrum and the symmetry is enhanced to $\su(2)_F \times \uu(1)$. This $\su(2)_F \times \uu(1)$ invariant $\mathcal{N}=2$ J-fold solutions was found also in five-dimensional gauged supergravity in \cite{Bobev:2020fon}.

We also find another distinct family of $\mathcal{N}=2$ J-fold solutions not discussed in \cite{Guarino:2020gfe} which we call Family II. It is invariant under an inequivalent $\Z_2$ subgroups of $S_4$, namely the $\langle f|f^2=e\rangle$. The 7 complex scalar fields along this family of AdS$_4$ solutions are
\be\label{eq:FamIIzs}
\text{Family II}:\quad{\bf z} = \left(\frac{i c
   \sqrt{\varphi^2+1}}{\sqrt{2}},\frac{i c
   \sqrt{\varphi^2+1}}{\sqrt{2}},i
   c,\frac{1+i}{\sqrt{2}},\frac{1+i}{\sqrt{2
   }},\frac{-\varphi+i}{\sqrt{\varphi^2+1}},\frac{\varphi+i}{\sqrt{\varphi^2+1}}\right)\,,
\ee
where the parameter $\varphi$ is a real scalar modulus with range $\varphi \in [0,\infty)$. The scalar potential for this family of solutions takes the value $V= -3 g^2/c$ which is the same as that of Family I. The spectrum of gravitino masses reads
\be\label{eq:gravmassesFamII}
m^2_{3/2}L^2~:~ 1_2\,,\quad 4_2 \,,\quad \left[\frac{(\varphi ^2-\varphi +2)^2}{2 (\varphi ^2+1)}\right]_2\,,\quad \left[\frac{(\varphi ^2+\varphi +2)^2}{2 (\varphi ^2+1)}\right]_2\,.
\ee
For generic values of $\varphi$ we have two massless vector fields in the supergravity spectrum and thus Family II also preserves $\uu(1) \times \uu(1)$ symmetry and $\mathcal{N}=2$ supersymmetry. Comparing \eqref{eq:FamIzs} and \eqref{eq:FamIIzs} along with the spectrum of supergravity fields we find that the points $\chi=0$ and $\varphi=0$ are equivalent and thus for $\varphi=0$ we recover the $\su(2)_F \times \uu(1)$ invariant $\mathcal{N}=2$ J-fold solution. The value $\varphi=1$ is also special since at this point there are 6 massless vector fields in the supergravity spectrum and the gravitino masses in \eqref{eq:gravmassesFamII} become $(1_4,4_4)$. Indeed, for $\varphi=1$ we find the $\mathcal{N}=4$ J-fold solution with $\su(2)\times \su(2)$ global symmetry studied in \cite{Inverso:2016eet,Guarino:2020gfe}. We note that the solution with $\varphi=1$ is also invariant under the $S_3$ subgroup of $S_4$ given by
\be
S_3 = \big \langle f,(hg),(fhg)|(hg)^2 = f^2 = (fhg)^3 = (hg)(f)(fhg)=e\rangle.
\ee
We discuss the full spectrum of the four-dimensional supergravity fields for Family II in Section~\ref{sec:spectroscopy}. Here we just note that Family I and II exhibit different degeneracies in the gravitino spectrum \eqref{eq:gravmassesFamI} and \eqref{eq:gravmassesFamII}. The extra degeneracy in Family I is not explained by the continuous $\uu(1) \times \uu(1)$ symmetry and is probably due to the discrete $\mathbf{Z}_2$ symmetry discussed above \eqref{eq:FamIzs}.

Our focus here is on studying AdS$_4$ solutions with $\mathcal{N}=2$ and we have not performed an exhaustive search for AdS$_4$ critical points of the scalar potential that are non-supersymmetric or have only $\mathcal{N}=1$ supersymmetry. There are certainly many such critical points of the scalar potential some of which were studied in \cite{Guarino:2020gfe}. For instance a family of non-supersymmetric solutions with $V=-2\sqrt{2} g^2/c$ was found in \cite{Guarino:2020gfe}. Curiously, we have found a large number of different nonsupersymmetric critical points of the 14-scalar model which have $V=-3 g^2/c$. This is  the same value of the cosmological constant as the ones for Family I and II above and there seems to be family of non-supersymmetric smoothly connected to the two supersymmetric families. These include for example the $\Z_2^2 \subset S_4$ symmetric point
\be
{\bf z} = \Big(\frac{i c}{\sqrt{3}},\frac{i c}{\sqrt{3}}
   ,i,i \sqrt[4]{3},i c
   \sqrt[4]{3},\frac{i}{\sqrt[4]{3}},\frac{i
   }{\sqrt[4]{3}}\Big)\,,
\ee
which enjoys an $\so(5)$ continuous symmetry. All gravitinos have normalized mass $m^2L^2=2$ at this AdS$_4$ vacuum and therefore no supersymmetry is preserved. Unfortunately, this AdS$_4$ is unstable since there are scalar fluctuations with $m^2L^2=-4$ which is below the BF bound. More generally, all non-supersymmetric AdS$_4$ critical points we found exhibit BF instabilities and we refrain from discussing them further here. It will be most interesting to perform a systematic and exhaustive search to find all AdS$_4$ vacuum solutions in the 14-scalar model along the lines of \cite{Bobev:2019dik,Bobev:2020qev} and study their stability.


\section{A two-parameter family of J-folds}
\label{sec:18}

Inspired by the two families presented in Section~\ref{sec:14}, and by our QFT discussion in Section~\ref{Sec: Introduction}, we expect a two-parameter family of J-fold solutions with $\mathcal{N}=2$ supersymmetry that interpolates between Families I and II. We quickly realize that this 2-parameter family of AdS$_4$ solutions is not inside the 14-scalar supergravity truncation discussed above. One way to see this is to study the spectrum of masses of the 14 scalars of the theory on a generic point on Family I or II and see that only one scalar is massless, i.e. there is only one candidate modulus which we denoted with $\chi$ on Family I and $\varphi$ on Family II. The problem we face is that the two families constructed above preserve a $\uu(1)_F \times \uu(1)_R$ symmetry that is embedded in different ways in the $\so(6)$ symmetry of the four-dimensional gauged supergravity. One can use $\so(6)$ group elements to rotate the solutions in Family I and II such that they preserve the same $\uu(1)^2$ symmetry, however this rotation involves turning on scalar fields that lie outside the 14-scalar truncation described above. This observation is precisely what allows us to construct a new two-parameter family of solutions as we now explain in more detail.

We first note that the solutions in Family I and II break some of the elements of the $S_4$ symmetry of the 14 scalar model \eqref{eq:S4symdef}. One can then rotate a given AdS$_4$ solution with such a broken $S_4$ element into a new ``gauge''. This means that the two families introduced above can be expressed in many equivalent ways by such $S_4$ rotations. Using this technique we can rotate both families of solutions such that they preserve a common $\uu(1)_R\subset \so(6)$ generator. Each family additionally preserves one more $\uu(1)$, but those are in general embedded differently inside $\so(6)$. We can act with $S_4$ generators on the solutions in Family I and II such that they share explicitly the same $\su(2)_F$ invariant point, i.e. the $\su(2)_F$ symmetries for the $\chi=0$ solution in Family I and the $\varphi=0$ solution of Family II are the same. The result of this operation is the following representation of Family I and II in a different gauge:
\be\label{newgaugeFamilies}
\begin{split}
\text{Family I}:&\quad {\bf z} = \Big( -c\chi+\frac{ic}{\sqrt{2}},-c\chi+\frac{ic}{\sqrt{2}},i c,i,i,\frac{1+i}{\sqrt{2}},\frac{1+i}{\sqrt{2}}\Big)\,,\\
\text{Family II}:&\quad{\bf z} = \Big(\frac{i c\sqrt{\varphi^2+1}}{\sqrt{2}},\frac{i c\sqrt{\varphi^2+1}}{\sqrt{2}},ic,\frac{-\varphi+i}{\sqrt{\varphi^2+1}},\frac{\varphi+i}{\sqrt{\varphi^2+1}},\frac{1+i}{\sqrt{2}},\frac{1+i}{\sqrt{2}}\Big)\,.
\end{split}
\ee
Notice that $\chi=0$ in the equation above gives the same values of $z_i$ as $\varphi=0$ and so by definition they preserve the same $\su(2)_F\subset \so(6)$. This is to be contrasted with the values of $z_i$ in \eqref{eq:FamIzs} and \eqref{eq:FamIIzs} which are permuted among each other at the $\chi=\varphi=0$ point.

For generic values of $\chi$ and $\varphi$ Family I and II preserve $\uu(1)^2$. The generator $\uu(1)$ that is preserved by both families of solutions is given by
\be
g_R=\f12\begin{bmatrix} 
 0 & 0 &0& 1 &0&0\\
0 & 0 &1& 0 &0&0\\
 0 & -1&0&0 &0&0\\
-1 & 0 &0& 0 &0&0\\
0  & 0 &0& 0 &0&0\\
0  & 0 &0& 0 &0&0
\end{bmatrix} \,,
\ee
where $g_R$ should be thought of as a generator inside $\so(6)$. This is dual to the superconformal $\uu(1)_R$ R-symmetry in the dual SCFT. The second $\uu(1)$ preserved by each family is dual to the flavor symmetry and is given by distinct $\so(6)$ generators. Using \eqref{newgaugeFamilies} we find that the generators take the following form
\be
g_\text{I}=\f12\begin{bmatrix} 
 0 & 1 &0& 0 &0&0\\
-1 & 0 &0& 0 &0&0\\
 0 & 0&0&-1 &0&0\\
0 & 0 &1& 0 &0&0\\
0  & 0 &0& 0 &0&0\\
0  & 0 &0& 0 &0&0
\end{bmatrix} \,,\qquad 
g_\text{II}=\f12\begin{bmatrix} 
 0 & 0 &1& 0 &0&0\\
0 & 0 &0& 1 &0&0\\
 -1 & 0&0&0 &0&0\\
0 & -1 &0& 0 &0&0\\
0  & 0 &0& 0 &0&0\\
0  & 0 &0& 0 &0&0
\end{bmatrix} \,.
\ee
We also find that $g_\text{I}$, $g_\text{II}$, and $[g_\text{I},g_\text{II}]$ are the three generators of $\su(2)_F$ at the point $\chi=\varphi=0$. We can now use the generator $[g_\text{I},g_\text{II}]$ to rotate one of the two solutions in \eqref{newgaugeFamilies} by the angle $\pi/2$ such that it preserves the same two $\uu(1)$ generators as the other. For example, we can do this to the Family I solutions explicitly at the level of the scalar coset element, or 70-bein ${\cal V}$. Recall that for the 14 scalar model the scalar coset is given by a product of $\SL(2,\R)/\U(1)$ elements in \eqref{70bein14scal}. We will use $\cal V_\text{I}$ and $\cal V_\text{II}$ to denote the 70-bein corresponding to the two solutions \eqref{newgaugeFamilies}. The vielbein $\cal V_\text{I}$ and $\cal V_\text{II}$ can be written as $56\times 56$ matrices that only depend on the moduli $\chi$ and $\varphi$, respectively. Using this notation we can rotate the Family I solutions, using the generator $[g_\text{I},g_\text{II}]$ as
\be
\tilde{\cal V}_\text{I} = \e^{[g_\text{I},g_\text{II}] \pi/2}\cdot{\cal V}_\text{I}\,.
\ee
Here the $\su(2)_F$ generator $[g_\text{I},g_\text{II}]$ should be embedded inside $\ess$ using the formulae in \cite{deWit:2007kvg}. After performing this rotation, the solution $\tilde{\cal V}_\text{I}$ no longer belongs to the 14-scalar supergravity truncation. Instead it is part of a larger $\uu(1)^2$ invariant truncation which is obtained by keeping only fields that are invariant under the generators $g_\text{II}$ and $g_R$ in $\so(6)$. There are 18 scalars fields in this bigger truncation which span the scalar manifold
\be\label{scalarmfd}
{\cal M} = \f{\SL(2,\R)}{\U(1)}\times \f{\SO(4,4)}{\SO(4)\times \SO(4)}\,.
\ee
We have explicitly parametrized this coset space and computed the scalar potential and superpotential as a function of the 18 scalars. The result is quite unwieldy and we refrain from presenting it here. Fortunately, we do not need the details of the action for this 18-scalar model to explain how to construct the two-parameter AdS$_4$ solutions of interest.

\begin{figure}[H]
\centering
\begin{overpic}[width=0.55\textwidth]{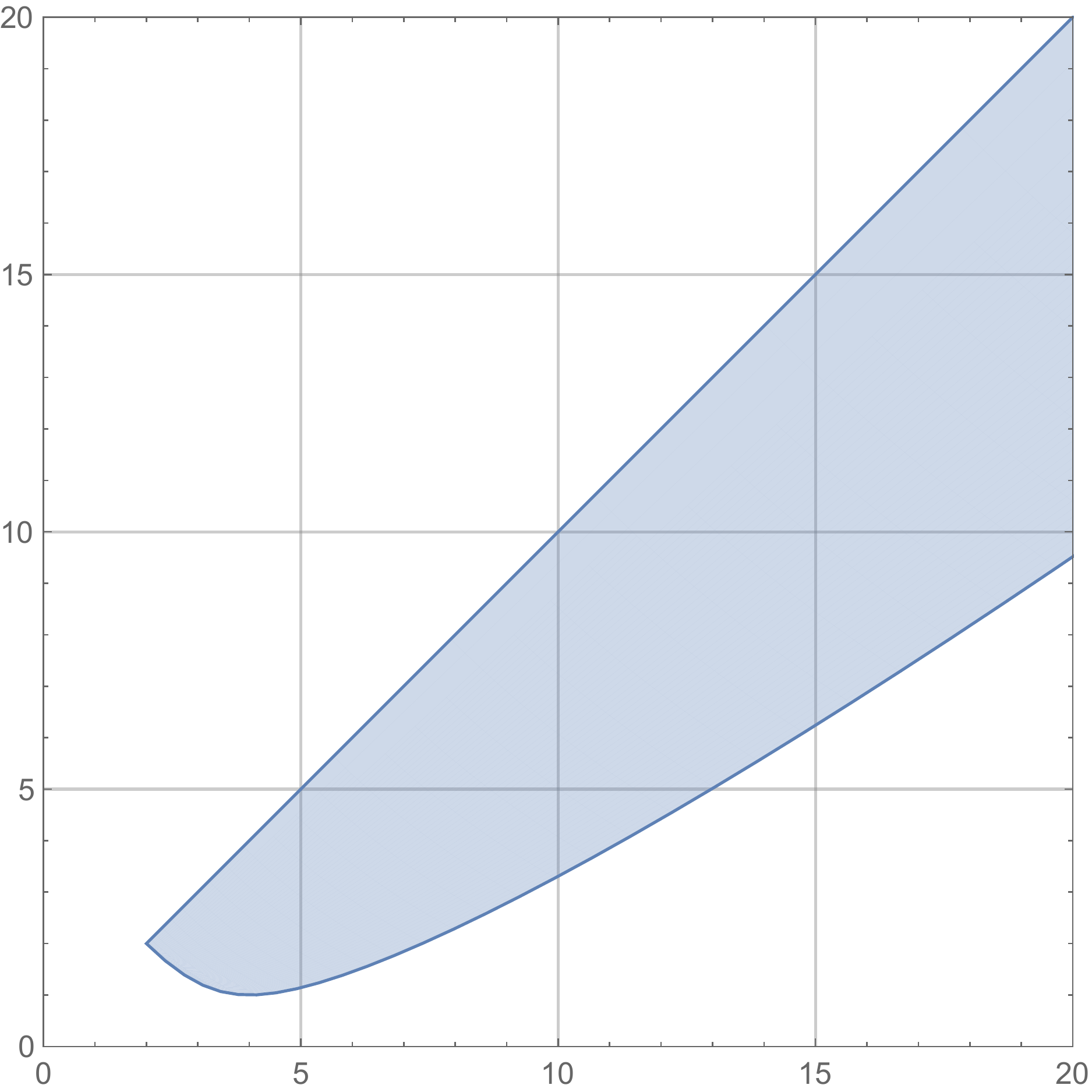}
\put (12.3,12.7){$\bullet$}
\put (5,16) {\footnotesize$\su(2)_F$}
\put (22,8){$\bullet$}
\put (18,5) {\footnotesize${\cal N}=4$}
\end{overpic}
\caption{\label{gmassplot} Parametric plot of the two non-trivial gravitino masses in \eqref{eq:gravmassgen} across the conformal manifold. The two curves that bound the blue region denote Family I and Family II. Family I corresponds to the upper straight line where enhanced degeneracies for the gravitino masses is observed. The lower curve represents the gravitino masses for Family II. The lowest point on the lower curve is located at gravitino masses $(4,1)$ where supersymmetry is enhanced to ${\cal N}=4$, while the leftmost solid point corresponds to the $\su(2)_F$ invariant $\mathcal{N}=2$ J-fold background. The shaded blue region depicts the allowed gravitino masses given by the analytic expressions in \eqref{eq:gravmassgen}. The conformal manifold appears to be non-compact but its global properties are subtle and will be discussed in Section~\ref{sec:discussion}.}
\end{figure}
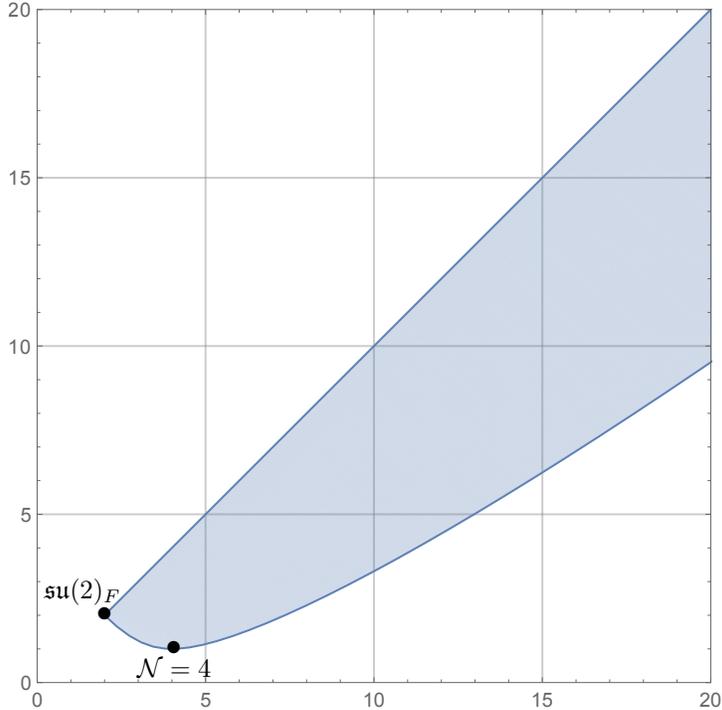

First we identify the element of the $\Ess/\SU(8)$ scalar coset which can be used to generate a Family I solution with arbitrary parameter $\chi$ by starting from the $\chi=0$ $\su(2)_F$ invariant solution with coset element ${\cal V}_{\su(2)_F}=\tilde{\cal V}_\text{I}|_{\chi=0}$. This coset element can be written as
\be
{\cal R}_\chi = \tilde{\cal V}_\text{I}\cdot ({\cal V}_{\su(2)_F})^{-1}\,.
\ee
We can then use ${\cal R}_\chi$ to write the scalar coset element for the full Family I as $\tilde{\cal V}_\text{I} = {\cal R}_\chi\cdot {\cal V}_{\su(2)_F}$. We can also use the matrix ${\cal R}_\chi$ to introduce the parameter $\chi$ all along Family II and obtain the two-parameter family of scalar coset elements
\be\label{themagicalsolution}
{\cal V}(\chi,\varphi) = {\cal R}_\chi\cdot {\cal V}_\text{II}\,.
\ee
Remarkably, one can show that the coset element ${\cal V}(\chi,\varphi)$ describes a two-parameter family of AdS$_4$ solutions of the scalar potential of the four-dimensional supergravity theory. We have confirmed this explicitly by computing the gradient of the scalar potential along all 70 directions and showing that it vanishes. For the entire family of solutions given by ${\cal V}(\chi,\varphi)$ the value of the scalar potential is $V=-3 g^2/c$ which is the same value as for Family I and II. Indeed, one can show that by setting $\varphi=0$ in \eqref{themagicalsolution} one recovers Family I and by setting $\chi=0$ one finds Family II. To establish that this two-parameter family of solutions is supersymmetric we can look at the gravitino masses which take the form 
\be\label{eq:gravmassgen}
m^2_{3/2}L^2~:~ 1_2\,,\quad 4_2 \,,\quad \left[\frac{\varphi^2+\left(\varphi ^2+2\right)^2+2 \chi ^2\pm 2 \varphi  \sqrt{\left(\varphi ^2+2\right)^2+2 \chi ^2}}{2 \left(1+\varphi
   ^2\right)}\right]_2\,.
\ee
We indeed find two gravitinos with $m^2L^2=1$ as expected for $\mathcal{N}=2$ AdS$_4$ vacua. The mass spectrum of all other supergravity fields in the theory will be discussed in detail in the next section. Here we only point out that for general values of the parameters $(\varphi,\chi)$ we have two massless vector fields in the spectrum, in harmony with the expected $\uu(1)_F\times \uu(1)_R$ symmetry. The non-trivial gravitino masses in \eqref{eq:gravmassgen} can be used as a parametrization of the two-dimensional space of AdS$_4$ vacua. This is illustrated in Figure~\ref{gmassplot} which makes it clear that the $\su(2)_F$ $\mathcal{N}=2$ and $\mathcal{N}=4$ J-fold solutions occupy special loci in the space of solutions. In particular, the gravitino masses are minimized at the $\mathcal{N}=4$ J-fold solution.

We close this section by discussing the kinetic terms for the two scalar moduli $\varphi$ and $\chi$. This can be obtained directly from the scalar vielbein in \eqref{themagicalsolution}. These kinetic terms define a natural metric on the space of AdS$_4$ solutions which is given by the line element
\be\label{eq:metchiphi}
ds^2_K=\f{1+2\varphi^2}{2(1+\varphi^2)^2}\Big( \dd\varphi^2 + 2(1+\varphi^2)\dd \chi^2 \Big)\,.
\ee
Note that $\partial_{\chi}$ is a Killing vector for the metric. The Ricci curvature of this two-dimensional space is 
\be\label{eq:RicZ}
\text{Ricci} = \f{4(4\varphi^4+2\varphi^2-1)}{(1+2\varphi^2)^3}\,.
\ee
We will come back to this metric, and its interpretation in the dual field theory in Section~\ref{sec:discussion}. Here we note that the Ricci scalar does not have a definite sign and approaches 0 for large $\varphi$. We also find a zero of the Ricci scalar at the positive real value $\varphi = \frac{1}{2}\sqrt{\sqrt{5}-1}$.

\section{Spectroscopy}
\label{sec:spectroscopy}

As discussed in Section~\ref{Sec: Introduction} the two-parameter family of J-fold backgrounds that we constructed above should be dual to a conformal manifold of 3d $\mathcal{N}=2$ SCFTs. These SCFTs are strongly interacting and holography offers a powerful tool to uncover their physics. In particular, the mass spectrum of supergravity and string theory excitations around a given AdS$_4$ background is mapped to the spectrum of conformal dimensions of dual SCFT operators. To this end we have computed the mass spectrum for all four-dimensional supergravity fields around the AdS$_4$ J-folds. These supergravity modes comprise the spin-2 graviton, 8 spin-3/2 gravitini, 28 spin-1 vector fields, 56 spin-1/2 gaugini as well as the 70 spin-0 scalar modes. To discuss the spectrum we employ the holographic dictionary to compute the conformal dimensions of the SCFT operators dual to each of these modes using the standard relations
\begin{equation}\label{Eq: holographic dictionary}
\begin{aligned}
 &[0]\,: \quad m^2 L^2 = \Delta (\Delta-3)\,,\\
&[\tfrac12]\,: \quad m L = \left(\Delta- \frac32\right)\,,\\
&[1]\,: \quad m^2 L^2 = (\Delta-2) (\Delta-1)\,,\\
&[\tfrac32]\,: \quad m L = \left(\Delta- \frac32\right)\,.
\end{aligned}
\end{equation}
Here $[j]$ denotes the Lorentz-spin of the CFT operator and $\Delta$ its the conformal dimension, $m$ is the mass of the supergravity mode, and $L$ is the AdS length scale. The graviton is not included in \eqref{Eq: holographic dictionary} since it is massless and dual to the $\Delta=3$ energy-momentum operator in the SCFT. Equipped with this data we can then organize the SCFT operators into ${\cal N}=2$ superconformal multiplets which allows for a compact presentation of the spectrum. We denote the ${\cal N}=2$ multiplets as
\begin{equation}
X \bar Y [\Delta ;j ; r ; F ]\,,
\end{equation}
where $\Delta$, $j$, and $r$ denote the conformal dimension, spin, and $\uu(1)_R$-charge of the superconformal primary of the multiplet $X\bar Y$ and $F$ is its $\uu(1)_F$ flavor charge. More details on the structure of ${\cal N}=2$ superconformal multiplets can be found in Appendix~\ref{App: d=3 multiplets}.

We find that along the entire conformal manifold the spectrum of CFT operators dual to supergravity fields in the 4D ${\cal N}=8$ theory can be arranged into short, semi-short, and long multiplets. The short and semi-short multiplets are protected, meaning that the conformal dimensions of the operators in the multiplet are independent of the position on the conformal manifold. At a generic point on the conformal manifold we find the following protected multiplets
\begin{equation}\label{Eq:shortsemishort}
\begin{aligned}
& A_1 \overline A_1 \left[ 2;1;0;0\right]\,,& \qquad   L\overline A_1\left[ \tfrac52;\tfrac12;+ 1;0\right]\,,\qquad   A_1 \overline L\left[ \tfrac52;\tfrac12;- 1;0\right]\,,\\
& A_2 \overline A_2\left[ 1;0;0;0 \right]\,,&  \qquad L\overline B_1 \left[ 2;0;+ 2;0 \right] \,,\hspace{0.15cm} \qquad B_1 \overline L \left[ 2;0;- 2;0 \right] \,.
\end{aligned}
\end{equation}
Here $A_1 \overline A_1$ is the $\mathcal N=2$ energy-momentum tensor multiplet, and $A_2 \overline A_2$ is the $\uu(1)_F$ flavor current multiplet. The remaining multiplets are semi-short including the $ L\overline B_1$ and $ B_1 \overline L$ which contain the two exactly marginal real operators.

The remaining operators are arranged in long multiplets. The conformal dimensions of the operators in these multiplets are not protected and depend on the position on the conformal manifold. We find the following eight long multiplets at a generic point on the two-dimensional manifold
\begin{equation}\label{Eq:longs}
\begin{aligned}
&L\overline L\left[\tfrac12+\beta_1; 0 ;0;\pm2 \right]\,, \qquad   L\overline L\left[\tfrac12+\beta_2;0 ;0;0\right]\,, \qquad  L\overline L\left[\tfrac12+\beta_3; 0 ;0;0\right]\,,\\
& L\overline L\left[\tfrac12+\beta_4; \tfrac12 ;0;\pm1 \right]\,, \qquad   L\overline L\left[\tfrac12+\beta_5; \tfrac12 ;0;\pm1 \right]\,.
\end{aligned}
\end{equation}
The five real functions $\beta_i$ depend on the two coordinates parametrizing the conformal manifold and determine the conformal dimensions of all operators in these long multiplets. When expressed in terms of the coordinates $(\varphi,\chi)$ they take the explicit form
\begin{equation}\label{eq:beta15}
\begin{split}
\beta_1^2 & =\frac{1}{4} + 2\varphi^2 + \frac{4\chi^2}{1+\varphi^2} \,,\\
\beta_2^2 &= \frac{17+\varphi^2}{4(1+\varphi^2)}\,, \qquad \beta_3^2= \frac{17+33\varphi^2}{4(1+\varphi^2)}\,,\\
\beta_4^2 &= \frac{\varphi^2+\left(\varphi ^2+2\right)^2+2 \chi ^2-2 \varphi  \sqrt{\left(\varphi ^2+2\right)^2+2 \chi ^2}}{2 \left(1+\varphi
   ^2\right)}\,, \\
\beta_5^2 &= \frac{\varphi^2+\left(\varphi ^2+2\right)^2+2 \chi ^2+2 \varphi  \sqrt{\left(\varphi ^2+2\right)^2+2 \chi ^2}}{2 \left(1+\varphi
   ^2\right)}\,.
\end{split}
\end{equation}
Note that $\beta_{2,3}$ depend only on the coordinate $\phi$ which implies that some long multiplets have conformal dimensions that stay constant as one varies $\chi$. It is instructive also to present the functions $\beta_i$ for the two special families of J-fold backgrounds. We find the following results
\begin{equation}\label{Eq:famIIbetas}
\begin{split}
\beta_i &\underset{\text{Family I}}{\longrightarrow}\left(\tfrac{\sqrt{1+ 16 \chi^2}}{2},\tfrac{\sqrt{17}}{2} , \tfrac{\sqrt{17}}{2} ,  \sqrt{2+\chi^2},\sqrt{2+\chi^2}\right)\,,\\
\beta_i &\underset{\text{Family II}}{\longrightarrow}\left( \frac{\sqrt{1+8\varphi^2}}{2}  , \frac{\sqrt{17+\varphi^2}}{2\sqrt{1+\varphi^2}} , \frac{\sqrt{17+33\varphi^2}}{2\sqrt{1+\varphi^2}} , \frac{2- \varphi + \varphi^2}{\sqrt{2(1+\varphi^2)}}, \frac{2+ \varphi + \varphi^2}{\sqrt{2(1+\varphi^2)}}\right)\,.
\end{split}
\end{equation}
We observe that Family I enjoys an additional degeneracy in the spectrum of operators. This is the unique one-dimensional submanifold of the conformal manifold that exhibts such a degeneracy. It is natural to conjecture that this degeneracy is due to the nature of the discrete flavor symmetry present for Family I, see the discussion above \eqref{eq:FamIzs}. Family II on the other hand does not exhibit any extra degeneracy and thus behaves like a generic one-dimensional subsapce of the conformal manifold.

There are two special points with enhanced symmetry on the conformal manifold at which the spectrum exhibits rearrangements. The point $(\varphi,\chi)=(0,0)$ is the $\mathcal{N}=2$ J-fold background with $\su(2)_F\times \uu(1)_R$ symmetry. At this point we find that the two long multiplets in \eqref{Eq:longs} with conformal dimensions determined by $\beta_1$ decomposes into the following short and semi-short multiplets:
\begin{equation}\label{Eq: scalar LL decomposition}
    L\overline L [1;0;0;\pm2] \rightarrow A_2 \overline A_2 [1;0;0;\pm2] + L\overline B_1 [2 ; 0;+2 ;\pm2]  + B_1\overline L [2 ; 0;-2 ;\pm2]\,.
\end{equation}
The two $A_2 \overline A_2$ multiplets combine with the $A_2 \overline A_2$ multiplet in \eqref{Eq:shortsemishort} to form the $\su(2)_F$ current multiplet. The $L\overline B_1$ and $B_1\overline L$ are charged under the  $\su(2)_F$ symmetry and, as discussed in \cite{Green:2010da}, should appear precisely at special points in the conformal manifold associated with continuous flavor symmetry enhancement. At $(\varphi,\chi)=(0,0)$ it is of course possible to reorganize all other superconformal multiplets into $\mathfrak{su}(2)_F$ representations. To this end we use the notation
\begin{equation}
    X \bar Y [\Delta ;j ; r  ]\otimes [\ell]\,,
\end{equation}
where $\ell$ denotes the $\mathfrak{su}(2)_F$ spin, and find that the spectrum at the point $(\varphi,\chi)=(0,0)$, including the multiplets presented in \eqref{Eq: scalar LL decomposition}, takes the form
\begin{equation}\label{Eq: spectrum icelandic bridge su2 point}
\begin{aligned}
& 2 \times L\overline L\left[\tfrac{1+\sqrt{17}}{2} ;0 ;0\right] \otimes \left[0\right] \,, \quad   2 \times L\overline L\left[\tfrac{1+2\sqrt{2}}{2} ;\tfrac12 ;0\right] \otimes \left[\tfrac12\right]\,,\\
&  L\overline A_1\left[ \tfrac52;\tfrac12;+ 1\right] \otimes \left[0\right] \,,\quad   A_1 \overline L\left[ \tfrac52;\tfrac12;- 1\right] \otimes \left[0\right] \,,\\
& L\overline B_1 \left[ 2;0;+ 2 \right] \otimes \left[1\right] \,,\hspace{0.15cm} \quad B_1 \overline L \left[ 2;0;- 2 \right] \otimes \left[1\right] \,,\\
& A_1 \overline A_1 \left[ 2;1;0\right]  \otimes \left[0\right]  \,,\qquad A_2 \overline A_2\left[ 1;0;0 \right] \otimes \left[1\right] \,,
\end{aligned}
\end{equation}
The $A_2 \overline A_2\left[ 1;0;0 \right] \otimes \left[1\right]$ multiplet is the short multiplet that contains the $\mathfrak{su}(2)_F$ currents.

The second special point is at $(\varphi,\chi)=(1,0)$ which corresponds to the $\mathcal{N}=4$ J-fold solution. At this point we find that the two long multiplets in \eqref{Eq:longs} with conformal dimensions controlled by $\beta_4$ decompose into the following short and semi-short multiplets
\begin{equation}
    L\overline L \left[\tfrac32;\tfrac12;0;\pm 1\right] \rightarrow A_1 \overline A_1 \left[\tfrac32;\tfrac12;0;\pm1\right] + L \overline A_2 \left[2;0;1;\pm1\right] + A_2 \overline L  \left[2;0;-1;\pm1\right]\,.
\end{equation}
The two $A_1 \overline A_1 \left[\tfrac32;\tfrac12;0;\pm1\right]$ multiplets above are short multiplets which contains supercurrents which signal the enhancement to $\mathcal{N}=4$ supersymmetry at this locus on the conformal manifold. Indeed, the full spectrum discussed above takes a special form at the point $(\varphi,\chi)=(1,0)$ and can be organized into two $\mathcal{N}=4$ superconformal multiplets.\footnote{See Appendix~\ref{App: d=3 multiplets} for more details on $\mathcal{N}=4$ superconformal representation theory.} The short $\mathcal N=2$ multiplets recombine into the $\mathcal N=4$ energy-momentum tensor multiplet as follows
\begin{equation}
  \left.\begin{array}{l}
        A_1 \overline A_1  [2;1;0;0]\\
         A_1 \overline A_1  [\tfrac32;\tfrac12;0;\pm1]\\
          A_2 \overline A_2  [1;0;0;0]
  \end{array}\right\} \rightarrow A_2[1;0;0,0]\,.
\end{equation}
The semi-short and long $\mathcal N=2$ multiplets recombine into a short $\mathcal N=4$ multiplet
\begin{equation}
  \left.\begin{array}{l}
        L \overline{L}  [3;0;0;0]\\
        L \overline L  [\tfrac52;\tfrac12;0;\pm1]\\
        L \overline L  [2;0;0;0]\\
        L \overline L  [2;0;0;\pm2]\\
        L \overline A_1  [\tfrac52;\tfrac12;1;0]   +  A_1 \overline L  [\tfrac52;\tfrac12;-1;0]\\
        L \overline A_2  [2;0;1;\pm1]  +  A_2 \overline L  [2;0;-1;\pm1]\\
        L \overline B_1  [2;0;2;0] +  B_1 \overline L  [2;0;-2;0] \\
  \end{array}\right\} \rightarrow B_2[2;0;1,1]\,.
\end{equation}
This concludes our analysis of the mass spectrum of four-dimensional supergravity fields and its implications for the operator spectrum along the conform manifold in the dual SCFT. In the next section we will discuss some additional aspects of the properties of the conformal manifold.

\section{Discussion}
\label{sec:discussion}

In this paper we constructed a two-parameter family of $\mathcal{N}=2$ AdS$_4$ backgrounds of four-dimensional maximal gauged supergravity with $[\SO(1,1)\times \SO(6)]\ltimes \mathbf{R}^{12}$ gauge group and a dyonic gauging. The generic solutions in this class preserve $\uu(1)_F\times \uu(1)_R$ symmetry and can be uplifted to J-fold backgrounds of type IIB string theory of the type discussed around \eqref{eq:Jfoldintro}. As outlined in some detail in Section~\ref{Sec: Introduction} these string theory vacua should be holographically dual to the conformal manifold of a class of 3d $\mathcal{N}=2$ SCFT obtained from the Gaiotto-Witten $T[\U(N)]$ by gauging its global symmetry and introducing a Chern-Simons term for the vector multiplet. This holographic duality passes some non-trivial consistency checks. First we note that the supersymmetry and the $\uu(1)_F\times \uu(1)_R$ symmetry of the AdS$_4$ solutions match nicely with those of the family of SCFTs discussed around \eqref{eq:WN2}. Moreover the one-dimensional complex conformal manifold parametrized by the superpotential coupling $\lambda$ in \eqref{eq:WN2} is mapped to the two-parameter space of vacua we find and on both manifolds there is a special point with $\mathcal{N}=4$ supersymmetry enhancement. We have also established that the mass spectrum of supergravity excitations around the whole family of AdS$_4$ solutions can be organized into $\mathcal{N}=2$ superconformal multiplets and exhibits the structure expected for an SCFT with a one-dimensional conformal manifold. Yet another non-trivial test of AdS/CFT is provide by comparing the $S^3$ free energy of the family of SCFTs with the AdS$_4$ on-shell action. For the $\mathcal{N}=4$ SCFT discussed around \eqref{eq:WN4} the $S^3$ free energy was computed by supersymmetric localization in the large $N$ limit in \cite{Assel:2018vtq} and reads
\begin{equation}\label{eq:FS3disc}
\mathcal{F}_{S^3} = \frac{N^2}{2} \text{arccosh}(k/2)\,.
\end{equation}
Here $N$ is the rank of the gauge group and $k$ is the integer Chern-Simons level. As discussed in \cite{Bobev:2020fon} this $S^3$ free energy is not modified by exactly marginal couplings, or equivalently, it is not affected by introducing the mass term $m^2\Phi^2$ discussed around \eqref{eq:WN2}. We therefore conclude that along the whole 3d $\mathcal{N}=2$ conformal manifold described by \eqref{eq:WN2} the $S^3$ free energy takes the form \eqref{eq:FS3disc}. As shown in \cite{Assel:2018vtq} and \cite{Bobev:2020fon} the supersymmetric localization result in \eqref{eq:FS3disc} agrees precisely with the AdS$_4$ on-shell action for the $\mathcal{N}=4$ and $\mathcal{N}=2$ $\su(2)_F\times \uu(1)_R$ invariant J-fold solutions, respectively. In type IIB string theory, $N$ is the number of D3-branes and $k$ determines the $\SL(2,\Z)_{\text{IIB}}$ matrix used in the J-fold identification. The two-parameter family of $\mathcal{N}=2$ AdS$_4$ backgrounds we found above all have the same value for the cosmological constant, $V = -3g^2/c$, as the $\mathcal{N}=4$ and $\mathcal{N}=2$ $\su(2)_F\times \uu(1)_R$ invariant J-fold solutions. This in turn translates into the same value of the on-shell action and we therefore find that for the entire family of supergravity solutions and for the full SCFT conformal manifold the $S^3$ free energy is given by \eqref{eq:FS3disc}.

\begin{figure}[H]
\centering
\begin{overpic}[width=0.55\textwidth]{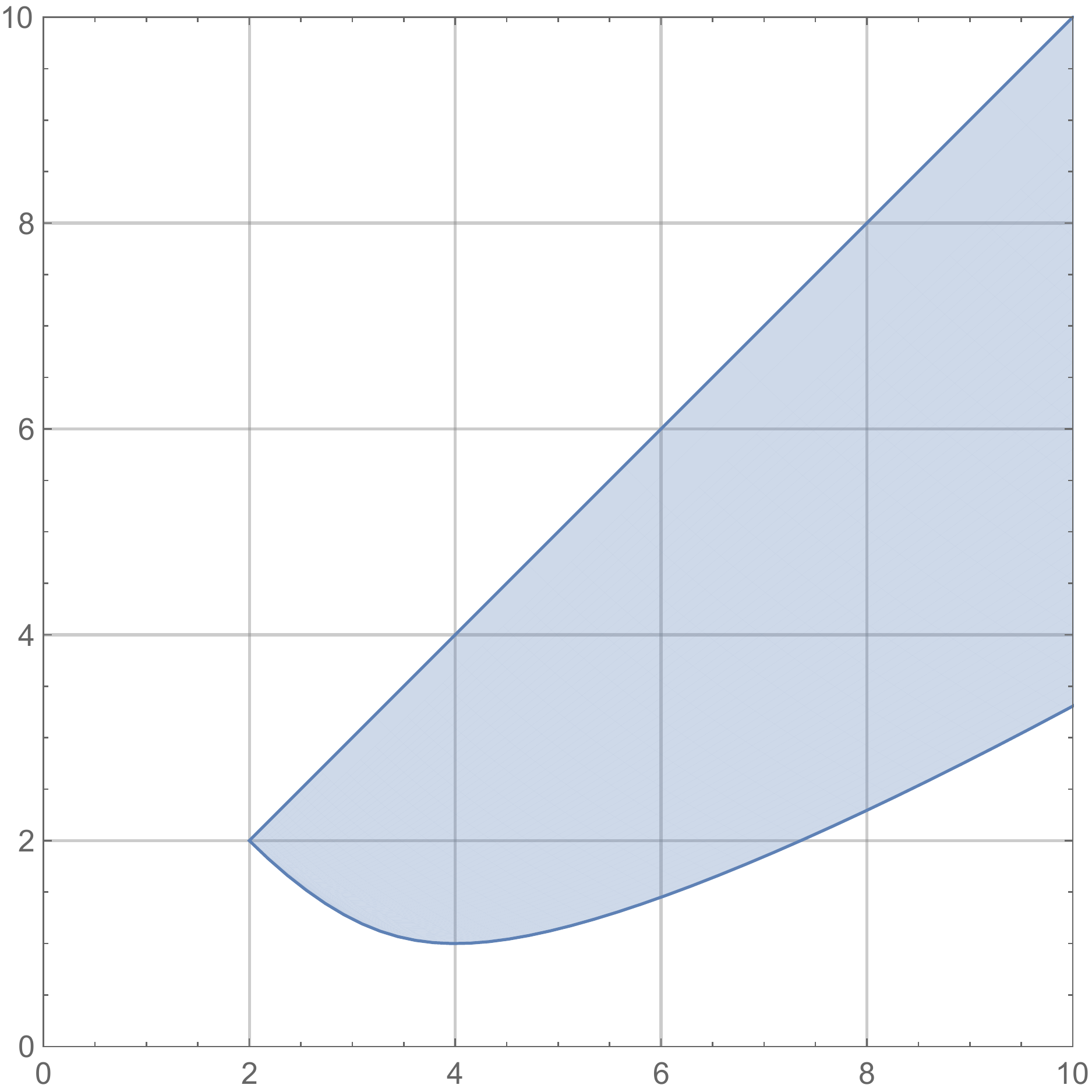}
\put (21.8,22.2){\DarkGreen $\bullet$}
\put (11,21) {\footnotesize$\su(2)_F$}
\put (41.8,42.2){\color{yellow}{$\bullet$}}
\put (27,46) {\footnotesize Ref. \cite{Giambrone:2021zvp}}
\put (61.8,62.2){\DarkGreen $\bullet$}
\put (81.8,82.2){\color{yellow}{$\bullet$}}
\put (40.5,12.3){\Red $\bullet$}
\put (36,8) {\footnotesize${\cal N}=4$}
\end{overpic}
\caption{\label{sketch} The known special points on $\mathcal{S}$. The green dots are different copies of the $\su(2)_F\times \uu(1)_R$ $\mathcal{N}=2$ J-fold background, the red dot is the $\mathcal{N}=4$ J-fold, and the yellow dots represent different copies of the AdS$_4$ J-fold solution with KK spectrum degeneracies discussed in \cite{Giambrone:2021zvp}.}
\end{figure}
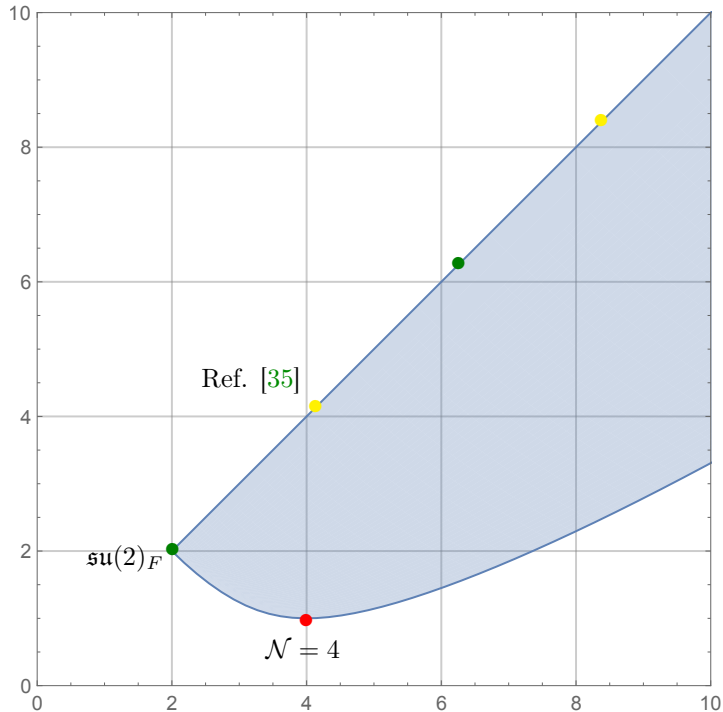

As discussed in Section~\ref{sec:14} the range of the scalar moduli parametrizing the space of AdS$_4$ J-fold backgrounds is $\varphi \in [0,\infty)$, $\chi \in [0,\infty)$.\footnote{These ranges can be extended to $\varphi \in (-\infty,\infty)$, $\chi \in (-\infty,\infty)$ but there is a symmetry of the solution space under $\varphi \to -\varphi$ and $\chi \to -\chi$ so we can restrict to the range $[0,\infty)$ without loss of generality.} This naively suggests that the conformal manifold in the dual three-dimensional $\mathcal{N}=2$ SCFT is non-compact. It is generally expected that non-compact conformal manifolds have loci where the SCFT at hand reduces to a free theory, see \cite{Perlmutter:2020buo} for a recent discussion. Based on this line of reasoning we therefore may conclude that the  $\mathcal{N}=2$ SCFTs described around \eqref{eq:WN2} admit a free limit for some value of the parameter~$\lambda$. While this is a logical possibility, it is not clear how this free SCFT arises from the strongly coupled non-Lagrangian $T[\U(N)]$ theory used in the construction. Moreover, if there is indeed a free SCFT somewhere along the conformal manifold we need to see hints of this in the SCFT spectrum we computed in Section~\ref{sec:spectroscopy}. The expectation is that at a putative free point, all operators should have half-integer conformal dimensions. From the results in \eqref{Eq:longs} and \eqref{eq:beta15} it is clear that we find half-integer conformal dimensions only at the $\mathcal{N}=4$ J-fold point which is certainly not expected to be dual to a free SCFT. One way to resolve this conundrum is to suppose that the space of AdS$_4$ vacua $\mathcal{S}$ parametrized by $(\varphi,\chi)$ is the covering space of the SCFT conformal manifold. The fundamental domain parametrizing inequivalent SCFTs, $\mathcal{M}_{\rm F}$, is then obtained by modding out $\mathcal{S}$ by some discrete group $\Gamma$, i.e. $\mathcal{M}_{\rm F}=\mathcal{S}/\Gamma$. Finding such equivalence relations on the space $\mathcal{S}$ using only four-dimensional supergravity does not seem feasible. Strong evidence for the existence of a non-trivial $\Gamma$ acting on $\mathcal{S}$ was found very recently in \cite{Giambrone:2021zvp} by studying the full type IIB supergravity spectrum of KK modes along the one-dimensional subspace of $\mathcal{S}$ given by Family I. As one increases $\chi$ the masses of many of the four-dimensional supergravity modes increase monotonically. At the same time however, the masses of some KK modes on the $\tilde{S}^1\times\hat{S}^5$ internal space decrease and the spectrum exhibits level crossing at some finite value of $\chi$. It was found in \cite{Giambrone:2021zvp} that the full KK spectrum has a periodicity and is equivalent for values of $\chi$ related by $\chi \to \chi + n\chi_0$, where $n$ is a positive integer and $\chi_0$ is determined by the Chern-Simons level $k$ in \eqref{eq:FS3disc}. This non-trivial spectrum rearrangement means that there are infinitely many copies of the $\mathcal{N}=2$ $\su(2)_F\times \uu(1)_R$ J-fold located at $n\chi_0$ along the one-dimensional space parametrized by $\chi$. It was also noted in \cite{Giambrone:2021zvp} that at the special values $\frac{n}{2}\chi_0$ the KK spectrum exhibits accidental degeneracies and the spectrum in the range $(\frac{n}{2}\chi_0,\frac{n+1}{2}\chi_0)$ appears to be identical to the one in the range $(\frac{n+1}{2}\chi_0,\frac{n+2}{2}\chi_0)$. This non-trivial structure of the KK spectrum naturally suggests that along Family I the inequivalent AdS$_4$ solutions lie in the range $(0,\frac{1}{2}\chi_0)$ and therefore the fundamental domain of the conformal manifold in the dual SCFT is obtained by some non-trivial action on $\mathcal{S}$. In Figure~\ref{sketch} we illustrate this structure by denoting the $\mathcal{N}=2$ $\su(2)_F\times \uu(1)_R$ J-folds with a green dot and the  special values $\frac{n}{2}\chi_0$ found in \cite{Giambrone:2021zvp} with a yellow dot.
 
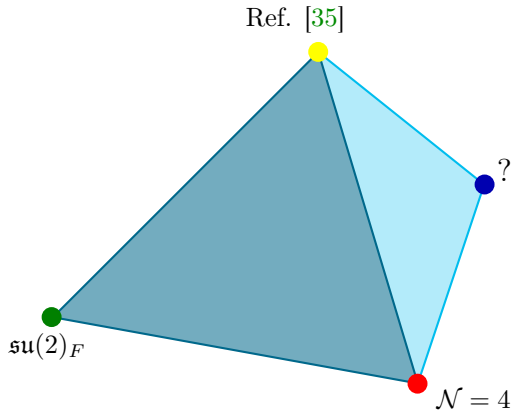
\begin{figure}[]
\centering
\begin{overpic}[width=0.35\textwidth]{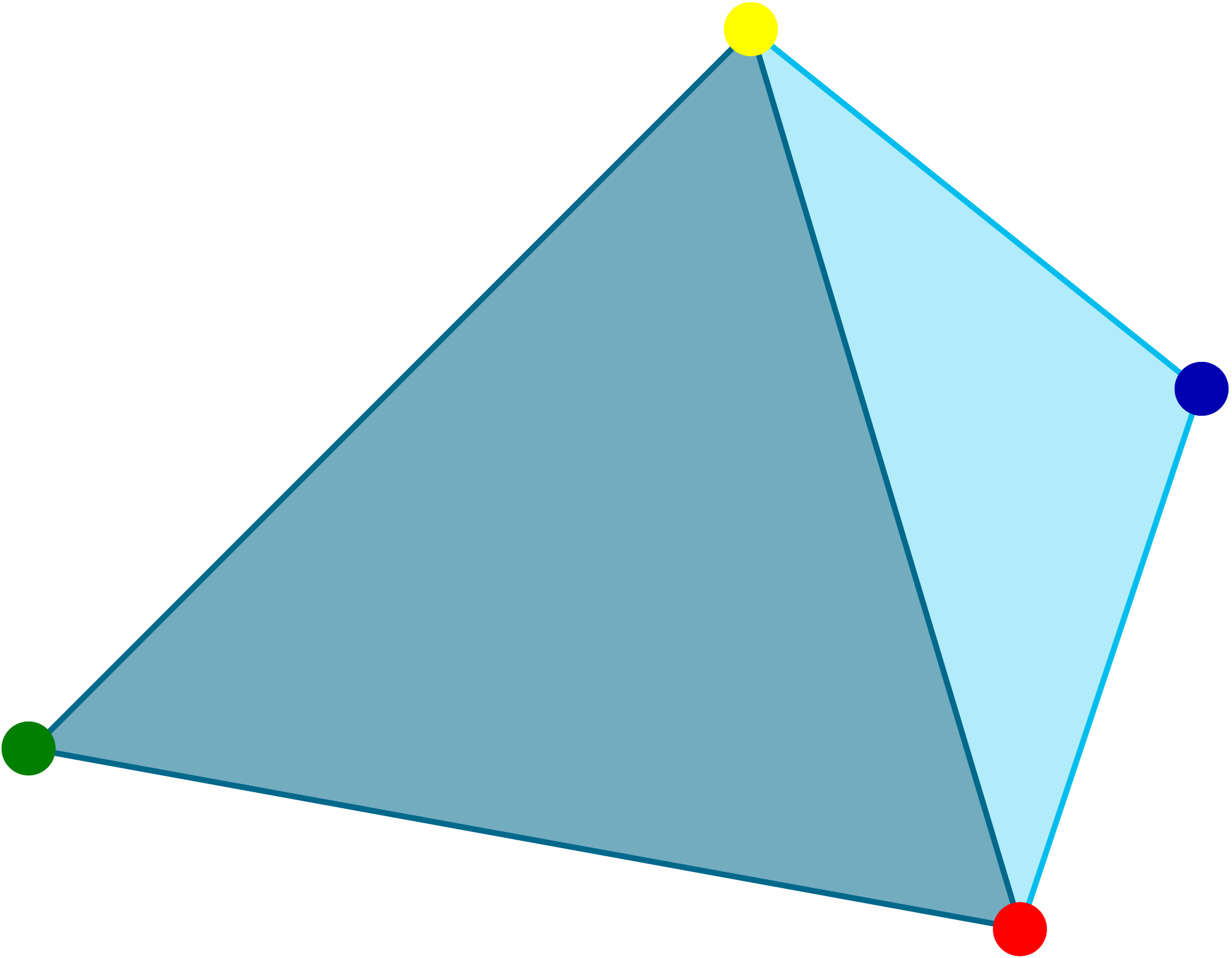}
\put (-7,9) {\footnotesize$\su(2)_F$}
\put (45,81) {\footnotesize Ref. \cite{Giambrone:2021zvp}}
\put (87,-3) {\footnotesize${\cal N}=4$}
\put (98,47) { ?}
\end{overpic}
\caption{\label{sketchtriangle} A sketch for the possible fundamental domain on the conformal manifold $\mathcal{M}_{\rm F}=\mathcal{S}/\Gamma$.}
\end{figure}

Based on the results of \cite{Giambrone:2021zvp} and our discussion above, it seems that to uncover the full structure of $\mathcal{M}_{\rm F}=\mathcal{S}/\Gamma$ using holography, one needs to uplift the family of solutions we found to ten dimensions and study the full KK spectrum of type IIB supergravity for the space $\mathcal{S}$.\footnote{The spin-2 KK spectrum of the $\mathcal{N}=4$ J-fold background was computed in \cite{Dimmitt:2019qla}.} In the absence of this explicit calculation we would like to offer a few comments on the possible structure of $\mathcal{M}_{\rm F}$. We have clear evidence from supergravity that there are three special points on $\mathcal{S}$ depicted by the green, yellow, and red dots in Figure~\ref{sketch}. This suggests that the fundamental domain $\mathcal{M}_{\rm F}$ may have a triangular shape as in the dark blue region of Figure~\ref{sketchtriangle}. It is also possible however that there are additional special points on $\mathcal{S}$. We have not seen evidence of any symmetry enhancement or spectrum degeneracy in our four-dimensional supergravity analysis in Section~\ref{sec:spectroscopy} that indicate the presence of such a point. However, it is entirely possible that the full KK spectrum along $\mathcal{S}$ exhibits special features similar to the one uncovered in \cite{Giambrone:2021zvp}. This will in turn lead to natural candidates for other special points that can act as additional vertices for the fundamental domain $\mathcal{M}_{\rm F}$. We have sketched this possibility in Figure~\ref{sketchtriangle} with a putative fourth special point leading to a quadrilateral shape for $\mathcal{M}_{\rm F}$. Settling the question about the full structure of $\mathcal{M}_{\rm F}$ conclusively requires uplifting the  two-parameter family of J-folds we found and calculating the full type IIB KK spectrum  and we hope to pursue this calculation in the future. In addition to helping uncover the structure of $\mathcal{M}_{\rm F}$, the explicit calculation of the KK spectrum will lead to detailed information about the spectrum of unprotected operators along the conformal manifold underscoring the unique utility of holography for addressing hard CFT questions. We note in passing that a similar structure to the one proposed above is exhibited by the conformal manifold of the three-dimensional $\mathcal{N}=2$ $XYZ$ model. As shown in \cite{Baggio:2017mas} the conformal manifold of this strongly interacting SCFT is an orbifold of $\mathbf{CP}^1$ with isolated fixed points and the fundamental domain has a triangular shape similar to the dark blue region of Figure~\ref{sketchtriangle}.

%

The kinetic terms for the scalar moduli $(\varphi,\chi)$ lead to the holographic result for the Zamolodchikov metric on the conformal manifold presented in \eqref{eq:metchiphi}. Such explicit results for the Zamolodchikov metric are rare and are of interest in their own right. For instance, the Zamolodchikov metric can be used to study the CFT distance conjecture discussed in \cite{Perlmutter:2020buo}. To this end we have analytically solved the geodesic problem for the metric in \eqref{eq:metchiphi} and have found that there are both bounded and unbounded geodesic curves. These results are of course valid on the space $\mathcal{S}$ and should be interpreted carefully when applied to the fundamental domain $\mathcal{M}_{\rm F}$ discussed above. Understanding this in detail is an interesting open problem that we intend to pursue in the future. We hope that this will also offer some insights into the peculiar structure of the Ricci scalar in \eqref{eq:RicZ} which does not have a definite sign on $\mathcal{S}$ and vanishes at $(\varphi,\chi) \to \infty$.

There are a number of possible generalizations and extensions of our work. It is clear that the four-dimensional maximal $[\SO(1,1)\times \SO(6)]\ltimes \mathbf{R}^{12}$ gauged supergravity has many interesting AdS$_4$ vacua. It will be most interesting to systematically classify these solutions by using the analytical and numerical tools developed to search for AdS vacua of other four-dimensional maximal supergravity theories \cite{Comsa:2019rcz,Bobev:2019dik,Bobev:2020qev}. It is notable that the structure of the potential of the $[\SO(1,1)\times \SO(6)]\ltimes \mathbf{R}^{12}$ gauged supergravity appears to be qualitatively different from that of other gauged supergravity theories with string and M-theory embeddings. Namely, as exhibited above, there are continuous families of stable AdS$_4$ vacua in this theory with different amount of supersymmetry. It will be particularly interesting to study families of $\mathcal{N}=1$ supersymmetric vacua like the examples constructed in \cite{Guarino:2020gfe}. These solutions should be holographically dual to conformal manifolds of three-dimensional $\mathcal{N}=1$ SCFTs which are hard to find and explore by purely field theory methods. A preliminary search indicates that the 18-scalar model \eqref{scalarmfd} admits new $\mathcal{N}=1$ AdS$_4$ solutions and it is desirable to study them further.

The type IIB supergravity uplifts of the class of J-fold backgrounds discussed here do not have singularities on the internal $\hat{S^5}$. There should be a vast generalization of these solutions where one allows for singular D5 and NS5 brane sources on $\hat{S}^5$ compatible with supersymmetry. Such generalizations of the $\mathcal{N}=4$ J-fold solution were studied in \cite{Assel:2018vtq} where they were shown to be holographically dual to quiver gauge theories significantly more involved than the simple SCFT discussed around \eqref{eq:WN4}. It will be most interesting to study $\mathcal{N}=2$ generalizations of this class of SCFTs and their dual AdS$_4$ string theory backgrounds. To this end one should probably extend the classification of AdS$_4$ $\mathcal{N}=4$ solutions of IIB supergravity discussed in \cite{DHoker:2007zhm,DHoker:2007hhe} to solutions with only $\mathcal{N}=2$. This appears to be a complicated task and it may be instructive to first look for specific examples of such $\mathcal{N}=2$ AdS$_4$ vacua in IIB supergravity guided by the uplift of the family of solutions discussed in this work.


\bigskip

\noindent \textbf{ Acknowledgements }

\medskip

\noindent We are grateful to Noppadol Mekareeya, Krzysztof Pilch, and Silviu Pufu for useful discussions. The work of NB is supported in part by an Odysseus grant G0F9516N from the FWO. FFG is supported by the University of Iceland Recruitment Fund. The work of JvM is supported by a doctoral fellowship from the Research Foundation - Flanders (FWO). NB and JvM are also supported by the KU Leuven C1 grant ZKD1118 C16/16/005.

\begin{appendices}



\section{Three-dimensional superconformal multiplets}
\label{App: d=3 multiplets}

In this appendix we summarize some facts about the representation theory for three-dimensional superconformal algebras. We follow the notation and conventions  in \cite{Cordova:2016emh}. The superconformal algebra in three dimensions is given by $\mathfrak{osp}(4|\mathcal N)$, where $\mathcal N$ denotes the number of supercharges. Here we are interested in discussing $\mathcal N=2$ and $\mathcal N=4$ SCFTs.

We start with the representation theory for $\mathcal N=2$. To label the states in the SCFT we use the notation $ [\Delta ;j ; r ]$, where $\Delta$ is the conformal dimension, $j$ denotes the $\mathfrak{su}(2)$ Lorentz spin, and $r$ is the charge under the $\mathfrak{u}(1)$ R-symmetry. There are four real supercharges with the following quantum numbers
\begin{equation}
Q \sim [\tfrac12 ;\tfrac12 ; -1 ] , \qquad  \overline{Q} \sim [\tfrac12 ;\tfrac12 ; +1 ] \,.
\end{equation}

Unitarity imposes bounds on the quantum numbers of SCFT operators. Generic unitarity multiplets obey these bounds and are usually referred to as ``long''. When the unitarity bounds are saturated the multiplets contain less states and are called ``short''. There are various possible shortening conditions that give rise to different superconformal multiplets. Since the shortening conditions are determined by $Q$ and or $\overline Q$ the corresponding multiplets can be categorized by two labels. To this end we use the following notation 
\begin{equation}
X \bar Y [\Delta ;j ; r ]\,.
\end{equation}
For $\mathcal N=2$ there are eight distinct kinds of multiplets listed below.
\begin{itemize}
\item long multiplets, $L\bar L\left[ \Delta; j; r \right]$, for which there is no shortening and one has the bound, 
\begin{equation}\label{eq:LLdef}
\Delta > 1 + j + |r|\,.
\end{equation}
\item semi-short multiplets:
\begin{itemize}
\item[{\tiny $\blacksquare$}] $L\bar A_1\left[ \Delta; j \geq \tfrac12 ;r>0 \right]$ and $A_1 \bar L\left[ \Delta; j\geq \tfrac12 ;r<0 \right]$  with
\begin{equation}
\begin{aligned}
L\bar A_1:\quad \Delta = 1+j+r\,,\qquad A_1 \bar L:\quad \Delta =1+j-r\,.
\end{aligned}
\end{equation}
\item[{\tiny $\blacksquare$}] $L\bar A_2\left[ \Delta;0 ;r>0 \right]$ and $A_2 \bar L\left[ \Delta;0;r<0 \right]$  with
\begin{equation}
\begin{aligned}
L\bar A_2:\quad &\Delta =1+r \,,\qquad A_2 \bar L: \quad&\Delta = 1  - r\,.
\end{aligned}
\end{equation}
\item[{\tiny $\blacksquare$}] $L\bar B_1\left[ \Delta;0;r>\tfrac12 \right]$, and $B_1 \bar L\left[ \Delta;0;r<-\tfrac12 \right]$  with
\begin{equation}
\begin{aligned}
L\bar B_1: \quad \Delta = r \,,\qquad B_1 \bar L: \quad  \Delta =  - r \,.
\end{aligned}
\end{equation}
\end{itemize}
\item short multiplets:
\begin{itemize}
\item[{\tiny $\blacksquare$}] $A_1 \bar A_1\left[ \Delta; j\geq \tfrac12 ; r=0 \right]$ with
\begin{equation}
\Delta = 1+j\,.
\end{equation}
\item[{\tiny $\blacksquare$}] $A_2 \bar A_2\left[ \Delta;j= 0; r=0\right]$ with
\begin{equation}
\Delta = 1\,.
\end{equation}
\item[{\tiny $\blacksquare$}] $A_2 \bar B_1\left[ \Delta; j=0;r=\tfrac12 \right]$ and $B_1 \bar A_2\left[ \Delta;j = 0 ;r=-\tfrac12 \right]$  with
\begin{equation}
\begin{aligned}
A_2 \bar B_1: \quad &\Delta = \frac12 \,, \qquad  B_1 \bar A_2: \quad \Delta = \frac12 \,.
\end{aligned}
\end{equation}
\item[{\tiny $\blacksquare$}] $B_1 \bar B_1\left[ \Delta; 0, 0 ; 0\right]$ is the identity operator with conformal dimension 
\begin{equation}
\begin{aligned}
&\Delta = 0 \,, \quad \text{and} \quad r =0 \,.\\
\end{aligned}
\end{equation}
\end{itemize}
\end{itemize}
For theories with $\mathcal N=4$ supersymmetry the R-symmetry is $\mathfrak{su}(2)_{r_1} \times \mathfrak{su}(2)_{r_2}$. The representations can therefore be labelled by four quantum numbers $ [\Delta ;j ; r_1,r_2 ]$, where $r_i$ are the spin labels of the two $\mathfrak{su}(2)$'s. There are eight supercharges that transform as 
\begin{equation}
Q \sim [\tfrac12;\tfrac12;\tfrac12,\tfrac12]
\end{equation}
The long and short superconformal multiplets can be labelled with one letter and we use the notation
\begin{equation}
X [\Delta ;j ; r_1,r_2 ]\,.
\end{equation}
The shortening conditions result in four distinct kinds of multiplets
\begin{itemize}
\item long multiplets, $L\left[ \Delta; j; r_1,r_2 \right]$, 
\begin{equation}\label{eq:Ldef}
\Delta > 1 + j + r_1 + r_2\,.
\end{equation}
\item short multiplets:
\begin{itemize}
\item[{\tiny $\blacksquare$}] $A_1\left[ \Delta; j \geq \tfrac12 ;r_1,r_2 \right]$  with
\begin{equation}
\begin{aligned}
 \Delta =1 + j +r_1 + r_2 \,.
\end{aligned}
\end{equation}
\item[{\tiny $\blacksquare$}] $A_2\left[ \Delta;0 ; r_1,r_2 \right]$  with
\begin{equation}
\begin{aligned}
\Delta =1+ r_1 + r_2  \,.
\end{aligned}
\end{equation}
\item[{\tiny $\blacksquare$}] $ B_1\left[ \Delta;0;r_1,r_2 \right]$ with
\begin{equation}
\begin{aligned}
 \Delta = r_1 + r_2  \,,
\end{aligned}
\end{equation}
\end{itemize}
\end{itemize}
Below we give an explicit representation of all states in the $\mathcal{N}=2$ and $\mathcal{N}=4$ multiplets that appear in the discussion in the main text. The $\mathcal N=2$ multiplets are organized in diagrams according to the $Q$, and $\overline Q$ descendants, while the $\mathcal N=4$ multiplets are simply organized in tables according to the $Q$ descendants. In order to save space we do not indicate the conformal dimension of individual states in the diagrams and tables and use the notation $[j]^{(r)}$ and $[j]^{(r_1,r_2)}$ for $\mathcal{N}=2$ and $\mathcal{N}=4$, respectively.
\begin{equ}[H]
\begin{equation*}
\begin{aligned}
& \xymatrix @C=5.7pc @R=5.7pc @!0 @dr {
{[1]^{(0)}} \ar[r]|--{{~\bar Q~}} \ar[d]|--{~Q~} 
& {[\tfrac32]^{(+1)}}  \ar[d]|--{~Q~} \\
{[\tfrac32]^{(-1)}  } \ar[r]|--{{~\bar Q~}}
& {[2]^{(0)}} 
}
\end{aligned}
\quad
\begin{aligned}
& \xymatrix @C=5.7pc @R=5.7pc @!0 @dr {
{[0]^{(0)}} \ar[r]|--{{~\bar Q~}} \ar[d]|--{~Q~} 
& {[\tfrac12]^{(+1)}}  \ar[d]|--{~Q~} \\
{[\tfrac12]^{(-1)}  } \ar[r]|--{{~\bar Q~}}
& {[1]^{(0)}} 
}
\end{aligned}
\end{equation*}
\caption{The $A_1\bar A_1  \left[ 2;1 ;0 \right]$ and $A_2 \bar A_2  \left[ 1;0 ;0 \right]$ multiplets.}\label{}
\end{equ}

\begin{equ}[H]
\begin{equation*}
\begin{aligned}
& \xymatrix @C=5.7pc @R=5.7pc @!0 @dr {
{[0]^{(2)}} \ar[d]|--{~Q~} \\
{[\tfrac12]^{(1)}} \ar[d]|--{~Q~} \\
{[0]^{(0)} }
}
\end{aligned}
\qquad
\begin{aligned}
& \xymatrix @C=5.7pc @R=5.7pc @!0 @dr {
{[0]^{(-2)}} \ar[r]|--{{~\bar Q~}} 
& {[\tfrac12]^{(-1)}} \ar[r]|--{{~\bar Q~}} 
& {[0]^{(0)}}
}
\end{aligned}
\end{equation*}
\caption{The $L\bar B_1  \left[ 2;0 ;r \right]$ and $B_1 \bar L  \left[ 2;0 ;r \right]$ multiplets.}\label{}
\end{equ}

\begin{equ}[H]
\begin{equation*}
\begin{aligned}
& \xymatrix @C=5.7pc @R=5.7pc @!0 @dr {
{[\tfrac12]^{(1)}} \ar[r]|--{{~\bar Q~}} \ar[d]|--{~Q~} 
& {[1]^{(2)}}  \ar[d]|--{~Q~} \\
{[1]^{(0)}\oplus [0]^{(0)}} \ar[r]|--{{~\bar Q~}} \ar[d]|--{~Q~} 
& {[\tfrac32]^{(1)} \oplus [\tfrac12]^{(1)}}\ar[d]|--{~Q~}\\
{[\tfrac12]^{(-1)} } \ar[r]|--{{~\bar Q~}}
& {[1]^{(0)}} 
}
\end{aligned}
\quad
\begin{aligned}
& \xymatrix @C=5.7pc @R=5.7pc @!0 @dr {
{[\tfrac12]^{(-1)}} \ar[r]|--{{~\bar Q~}} \ar[d]|--{~Q~} 
& {[1]^{(0)}\oplus [0]^{0}} \ar[r]|--{{~\bar Q~}} \ar[d]|--{~Q~} 
& {[\tfrac12]^{(1)}} \ar[d]|--{~Q~}\\
{[1]^{(-2)}} \ar[r]|--{{~\bar Q~}}
& {[\tfrac32]^{(-1)}  \oplus  [\tfrac12]^{(-1)}}\ar[r]|--{{~\bar Q~}}
 & {[1]^{(0)}} 
}
\end{aligned}
\end{equation*}
\caption{The $L\bar A_1  \left[ \tfrac52;\tfrac12 ;1 \right]$ and $A_1 \bar L  \left[\tfrac52;\frac12  ;-1 \right]$ multiplets.}\label{}
\end{equ}

\begin{equ}[H]
\begin{equation*}
\begin{aligned}
& \xymatrix @C=5.7pc @R=5.7pc @!0 @dr {
{[0]^{(1)}} \ar[r]|--{{~\bar Q~}} \ar[d]|--{~Q~} 
& {[\tfrac12]^{(2)}}  \ar[d]|--{~Q~} \\
{[\tfrac12]^{(0)} } \ar[r]|--{{~\bar Q~}} \ar[d]|--{~Q~} 
& {[1]^{(1)} \oplus [0]^{(1)}}\ar[d]|--{~Q~}\\
{[0]^{(-1)} } \ar[r]|--{{~\bar Q~}}
& {[\tfrac12]^{(0)}} 
}
\end{aligned}
\quad
\begin{aligned}
& \xymatrix @C=5.7pc @R=5.7pc @!0 @dr {
{[0]^{(-1)}} \ar[r]|--{{~\bar Q~}} \ar[d]|--{~Q~} 
& {[\tfrac12]^{(0)} } \ar[r]|--{{~\bar Q~}} \ar[d]|--{~Q~} 
& {[0]^{(1)}} \ar[d]|--{~Q~}\\
{[1]^{(-2)}} \ar[r]|--{{~\bar Q~}}
& {[1]^{(-1)}  \oplus  [0]^{(-1)}}\ar[r]|--{{~\bar Q~}}
 & {[\tfrac12]^{(0)}} 
}
\end{aligned}
\end{equation*}
\caption{The $L\bar A_2  \left[ 2;0 ;1 \right]$ and $A_2 \bar L  \left[2; 0  ;-1 \right]$ multiplets.}\label{}
\end{equ}

\begin{equ}[H]
\begin{equation*}
\begin{aligned}
& \xymatrix @C=5.7pc @R=5.7pc @!0 @dr {
{[\tfrac12]^{(r)}} \ar[r]|--{{~\bar Q~}} \ar[d]|--{~Q~} 
& {[1]^{(r+1)}+ [0]^{(r+1)}} \ar[r]|--{{~\bar Q~}} \ar[d]|--{~Q~} 
& {[\tfrac12]^{(r+2)}} \ar[d]|--{~Q~}\\
{[1]^{(r-1)}+ [0]^{(r-1)}} \ar[r]|--{{~\bar Q~}} \ar[d]|--{~Q~} 
& {[\tfrac12]^{(r)}+ [\tfrac32]^{(r)}+ [\tfrac12]^{(r)}}\ar[r]|--{{~\bar Q~}}\ar[d]|--{~Q~}
 & {[1]^{(r+1)}+ [0]^{(r+1)}} \ar[d]|--{~Q~}\\
{[\tfrac12]^{(r-2)}} \ar[r]|--{{~\bar Q~}}
& {[1]^{(r-1)}+ [0]^{(r-1)}} \ar[r]|--{{~\bar Q~}}  
& {[\tfrac12]^{(r)}} 
}
\end{aligned}
\quad
\begin{aligned}
& \xymatrix @C=5.7pc @R=5.7pc @!0 @dr {
{[0]^{(r)}} \ar[r]|--{{~\bar Q~}} \ar[d]|--{~Q~} 
& { [\tfrac12]^{(r+1)} } \ar[r]|--{{~\bar Q~}} \ar[d]|--{~Q~} 
& { [0]^{(r+2)} } \ar[d]|--{~Q~}\\
{ [\tfrac12]^{(r-1)}} \ar[r]|--{{~\bar Q~}} \ar[d]|--{~Q~} 
& {[0]^{(r)} +[1]^{(r)} + [0]^{(r)}}\ar[r]|--{{~\bar Q~}}\ar[d]|--{~Q~}
 & {[\tfrac12]^{(r+1)}} \ar[d]|--{~Q~}\\
{[0]^{(r-2)}} \ar[r]|--{{~\bar Q~}}
& {[\tfrac12]^{(r-1)} } \ar[r]|--{{~\bar Q~}}  
& {[0]^{(r)}} 
}
\end{aligned}
\end{equation*}
\caption{The $L\bar L  \left[ \Delta;\frac12 ;r \right]$ and $L \bar L  \left[ \Delta;0 ;r \right]$ multiplets. }\label{}
\end{equ}

\begin{table}[H]
	\centering
\renewcommand{\arraystretch}{2}
\begin{tabular}{ccccc}
& $Q$ & $Q^2$ & $Q^3$ & $Q^4$ \\ \hline\hline 
$[0]^{(0,0)}$& $[\tfrac12]^{(1/2,1/2)}$ & $ [1]^{(1,0)}$ & $[\tfrac32]^{(1/2,1/2)}$ & $[2]^{(0,0)}$ \\ 
&  & $ [1]^{(0,1)}$ &  &  \\ 
&  & $[0]^{(0,0)}$ &  & \\\hline\hline
\end{tabular} 
\caption{The $A_2[1;0;0,0]$ multiplet.}\label{Tab: N=4 EM multiplet}
\end{table}
\begin{table}[H]
	\centering
\renewcommand{\arraystretch}{2}
\begin{tabular}{ccccccc}
& $Q$ & $Q^2$ & $Q^3$ & $Q^4$ & $Q^5$ & $Q^6$ \\ \hline
$\Delta = 2$ & $5/2$ & $3$ & $7/2$ & $4$ & $9/2$ & 5 \\\hline\hline
$[0]^{(1,1)}$ & $[\tfrac12]^{(3/2,1/2)}$ & $ [1]^{(1,0)}$ &$ [\tfrac32]^{(1/2,1/2)}$ &$  [1]^{(0,1)}$ &$  [\tfrac12]^{(1/2,1/2)} $&$ [0]^{(0,0)}$ \\ 
& $ [\tfrac12]^{(1/2,1/2)}$ & $ [1]^{(1,1)}$ & $ [\tfrac12]^{(3/2,1/2)}$ & $ [1]^{(0,0)}$ &  & \\ 
& $[\tfrac12]^{(1/2,3/2)}$ & $ [1]^{(0,1)}$ & $ [\tfrac12]^{(1/2,3/2)}$ & $ [1]^{(1,0)}$ &  & \\ 
&  & $[0]^{(2,0)}$ & $2 \times [\tfrac12]^{(1/2,1/2)}$ &$  [0]^{(1,1)}$ &  & \\ 
&  & $ [0]^{(1,1)}$ &  & $ [0]^{(0,0)}$ &  & \\
&  & $[0]^{(0,2)}$ &  &  &  & \\ 
&  & $[0]^{(1,0)}$ &  &  &  & \\ 
&  & $[0]^{(0,1)}$ &  &  &  & \\ 
&  & $[0]^{(0,0)}$ &  &  &  & \\\hline\hline
\end{tabular} 
\caption{The $B_1[2;0;1,1]$ multiplet.}\label{Tab: N=4 B1 multiplet}
\end{table}
%


\end{appendices}

\bibliography{SFoldCM}
\bibliographystyle{utphys}

\end{document}